\documentclass[twocolumn,amsmath,resetfootnote]{aastex702}

\graphicspath{{figures/}}

\begin{document}

\title{Thermal-pulse-resolved Binary Population Synthesis with Detailed Stellar Models: Implications for Long-period WD+MS Formation}

\author[orcid=0009-0001-2504-7054,gname=Yanxu,sname=Shi]{Yanxu Shi}
\affiliation{Yunnan Observatories, Chinese Academy of Sciences (CAS), Kunming 650216, People's Republic of China}
\affiliation{International Centre of Supernovae, Yunnan Key Laboratory of Supernova Research, Kunming 650216, People's Republic of China}
\affiliation{University of Chinese Academy of Sciences, Beijing 100049, People's Republic of China}
\email{2675223153@qq.com}

\author[orcid=0000-0002-6398-0195,gname=Hongwei,sname=Ge]{Hongwei Ge}
\affiliation{Yunnan Observatories, Chinese Academy of Sciences (CAS), Kunming 650216, People's Republic of China}
\affiliation{International Centre of Supernovae, Yunnan Key Laboratory of Supernova Research, Kunming 650216, People's Republic of China}
\affiliation{University of Chinese Academy of Sciences, Beijing 100049, People's Republic of China}
\email{gehw@ynao.ac.cn}

\author[orcid=0000-0003-1535-0866,gname=Diogo,sname=Belloni]{Diogo Belloni}
\affiliation{Yunnan Observatories, Chinese Academy of Sciences (CAS), Kunming 650216, People's Republic of China}
\affiliation{International Centre of Supernovae, Yunnan Key Laboratory of Supernova Research, Kunming 650216, People's Republic of China}
\email{.}

\author[orcid=0000-0003-4265-7783,gname=Dengkai,sname=Jiang]{Dengkai Jiang}
\affiliation{Yunnan Observatories, Chinese Academy of Sciences (CAS), Kunming 650216, People's Republic of China}
\affiliation{International Centre of Supernovae, Yunnan Key Laboratory of Supernova Research, Kunming 650216, People's Republic of China}
\affiliation{University of Chinese Academy of Sciences, Beijing 100049, People's Republic of China}
\email{dengkai@ynao.ac.cn}

\author[orcid=0009-0006-9211-2860,gname='Hailiang', sname='Chen']{Hailiang Chen}
\affiliation{Yunnan Observatories, Chinese Academy of Sciences (CAS), Kunming 650216, People's Republic of China}
\affiliation{International Centre of Supernovae, Yunnan Key Laboratory of Supernova Research, Kunming 650216, People's Republic of China}
\affiliation{University of Chinese Academy of Sciences, Beijing 100049, People's Republic of China}
\email{chenhl@ynao.ac.cn}

\author[orcid=0000-0001-9204-7778,gname=Zhanwen,sname=Han]{Zhanwen Han}
\affiliation{Yunnan Observatories, Chinese Academy of Sciences (CAS), Kunming 650216, People's Republic of China}
\affiliation{International Centre of Supernovae, Yunnan Key Laboratory of Supernova Research, Kunming 650216, People's Republic of China}
\affiliation{University of Chinese Academy of Sciences, Beijing 100049, People's Republic of China}
\email{.}

\correspondingauthor{Hongwei Ge}
\email{gehw@ynao.ac.cn}

\begin{abstract}
The origin of the long-orbital-period white dwarf--main-sequence (WD+MS) binaries revealed by Gaia challenges the traditional association of common-envelope evolution (CEE) with short orbital periods. We develop a thermal-pulse-resolved binary population synthesis framework based on 51 detailed MESA evolutionary tracks with initial masses of 1--6 $M_{\odot}$. Applying these tracks to $5\times10^6$ initially sampled binaries, we follow the time-dependent stellar radius, wind mass loss, angular momentum, and tidal interactions, while calculating the envelope binding energy directly from the donor structure at the onset of Roche-lobe overflow. We find that thermal pulses can trigger binary interaction through envelope expansion and tidal orbital shrinkage, while the weakly bound envelopes of highly evolved thermally pulsing asymptotic giant branch (TP-AGB) donors allow CE ejection with limited orbital contraction. For our standard model, with $\alpha_{\mathrm{CE}}$ = 0.3 and $\alpha_{\mathrm{th}}$ = 0.5, systems with orbital periods above 30 days constitute approximately 11.5\% of the resulting TP-AGB-initiated post-CE binaries (PCEBs). Long-period PCEBs can form across a large range of CE parameters and can reach periods of several hundred days even without any internal-energy contribution when $\alpha_{\mathrm{CE}}$ = 1. Their WD masses peak near 0.6 $M_\odot$, whereas systems containing more massive WDs are rare. Moreover, long-period PCEBs closely follow the single-star initial--final mass relation, providing a potential diagnostic of their formation channel in clusters, where progenitor masses can be independently constrained. Our results identify late TP-AGB CEE as a viable contributor to the observed long-period WD+MS population.
\end{abstract}


\section{Introduction} \label{sec:introduction}

White dwarfs (WDs) are the final evolutionary products of low- and intermediate-mass stars. Given the high occurrence of binary systems \citep{2013ARA&A..51..269D,2017ApJS..230...15M,2024PrPNP.13404083C}, binaries consisting of a WD and a main-sequence (MS) companion represent an important population of stellar remnants. These systems are potential progenitors of various exotic stars, including Type Ia supernovae \citep{1984ApJ...277..355W,2004MNRAS.350.1301H}, cataclysmic variables \citep{1987ApJ...314..653W,2003A&A...406..305S}, symbiotic stars \citep{1995ApJ...447..656Y}, AM CVn \citep{2003MNRAS.340.1214P}, and low-frequency gravitational wave sources \citep{2001A&A...375..890N}. Understanding their formation and properties is therefore essential for studying the evolution of interacting binary populations.

Traditionally, detached WD+MS binaries have been classified into three formation channels \citep{2004A&A...419.1057W}: common-envelope evolution (CEE, \citealt{1975PhDT.......165W,1976IAUS...73...75P}), stable mass transfer (MT), and non-interacting evolution. Dynamically unstable MT from a red giant branch (RGB) or asymptotic giant branch (AGB) donor initiates a CE phase, during which rapid inspiral substantially contracts the orbit, typically producing short-period post-CE binaries (PCEBs). Stable MT avoids this rapid inspiral and can leave substantially wider orbits. For low-mass donors with degenerate cores, the core mass--radius relation produces a characteristic WD mass--orbital period relation \citep{1995MNRAS.273..731R,1999A&A...350..928T,2014MNRAS.437.2217S,2026ApJ...996...74M}. Still wider binaries may avoid Roche-lobe overflow, allowing approximately single-star evolution. These channels were expected to produce relatively well-separated period distributions, with a sparsely populated ``period gap'' between close CEE products and wider stable-MT or non-interacting systems \citep{2004A&A...419.1057W,2019MNRAS.484.5362A,2022MNRAS.512.2625L}. Orbital period therefore became a primary formation-channel indicator, with short-period systems associated with CEE and longer-period systems with stable MT.

However, recent discoveries have fundamentally challenged this traditional picture. The Gaia mission has revealed a large population of detached WD+MS binaries with orbital periods of approximately 100--1000 days \citep{2024MNRAS.529.3729S,2024PASP..136h4202Y,2024MNRAS.52711719Y}, populating precisely the range previously expected to contain relatively few systems. These binaries have orbital periods much longer than those of classical PCEBs, yet many contain WDs more massive than 0.5 $M_{\odot}$, suggesting that their progenitors likely evolved through the AGB phase. Their relatively compact orbits suggest that many of their progenitors may have undergone MT during the giant phases, yet their present-day properties do not fit neatly into the conventional picture of either CEE or stable MT. Recent studies have demonstrated that unusually wide WD+MS binaries may form through CEE involving highly evolved AGB donors \citep{2024A&A...686A.226B,2024A&A...687A..12B,2024A&A...686A..61B,2024PASP..136h4202Y,2024MNRAS.52711719Y}, while stable MT from intermediate-mass donors can also produce systems outside the classical WD mass--period relation \citep{2026ApJ..1009...74Z}. Indeed, both population synthesis calculations and observational studies increasingly indicate that the orbital-period distributions resulting from different evolutionary channels can overlap substantially \citep{2025A&A...704A...6R,2026ApJS..286...13V}. Thus, orbital period alone can no longer be regarded as a reliable discriminator of the formation channel, challenging a long-standing assumption in the study of WD+MS binaries.

In single low- and intermediate-mass stars, the thermally pulsing AGB (TP-AGB) begins when the helium-burning shell becomes thermally unstable. During each interpulse interval, hydrogen-shell burning adds helium to the layer beneath it until the next helium-shell flash. The flash drives convection between the burning shells and causes the envelope to expand. Over successive cycles, shell burning can grow the core while strong stellar winds strip the envelope, eventually ending the AGB phase and leaving a WD remnant \citep{2014ApJ...782...17K}. The TP-AGB phase has two important consequences for binary evolution. Pulse-driven envelope expansion can trigger Roche-lobe overflow (RLOF) in otherwise detached binaries, while extensive wind mass loss reduces the envelope mass and can facilitate CE ejection. Three-dimensional hydrodynamic simulations show that pulse-driven expansion can initiate rapid CE inspiral \citep{2022MNRAS.517.3181G}, and detailed stellar and binary models show that highly evolved TP-AGB interactions can produce unusually wide PCEBs \citep{2024A&A...686A.226B,2024A&A...687A..12B,2024PASP..136h4202Y,2024MNRAS.52711719Y}. Thermal pulses can thus influence both interaction onset and the final post-CE orbit.

Despite these advances, the population-level consequences have not yet been systematically explored. Widely used rapid binary population synthesis (BPS) codes, including BSE \citep{2000MNRAS.315..543H,2002MNRAS.329..897H}, COSMIC \citep{2020ApJ...898...71B}, and COMPAS \citep{2022ApJS..258...34R}, use analytic stellar-evolution prescriptions that do not resolve individual thermal pulses. Synthetic TP-AGB models in the \texttt{binary\_c} code \citep{2004MNRAS.350..407I} follow pulse cycles but do not provide detailed stellar structures during helium-shell flashes. Incorporating thermal pulses into BPS requires more than an improved binding-energy prescription: changes in radius, wind mass loss, and tidal interactions jointly determine the orbit and onset of RLOF. Because CE outcomes depend on the donor structure and orbit at interaction, these effects must be followed together throughout the pre-interaction evolution.

In this work, we incorporate detailed stellar models from the zero-age main sequence (ZAMS) through the TP-AGB into a pulse-resolved BPS framework. Following the stellar radius, mass loss, envelope binding energy, moment of inertia, and tidal interactions, we investigate how thermal pulses shape the formation and properties of long-period WD+MS binary population formed through CEE.

\section{Detailed Stellar Evolution Grid} \label{sec:stellar_grid}

\begin{figure*}[ht!]
\centering
\includegraphics[width=0.8\textwidth]{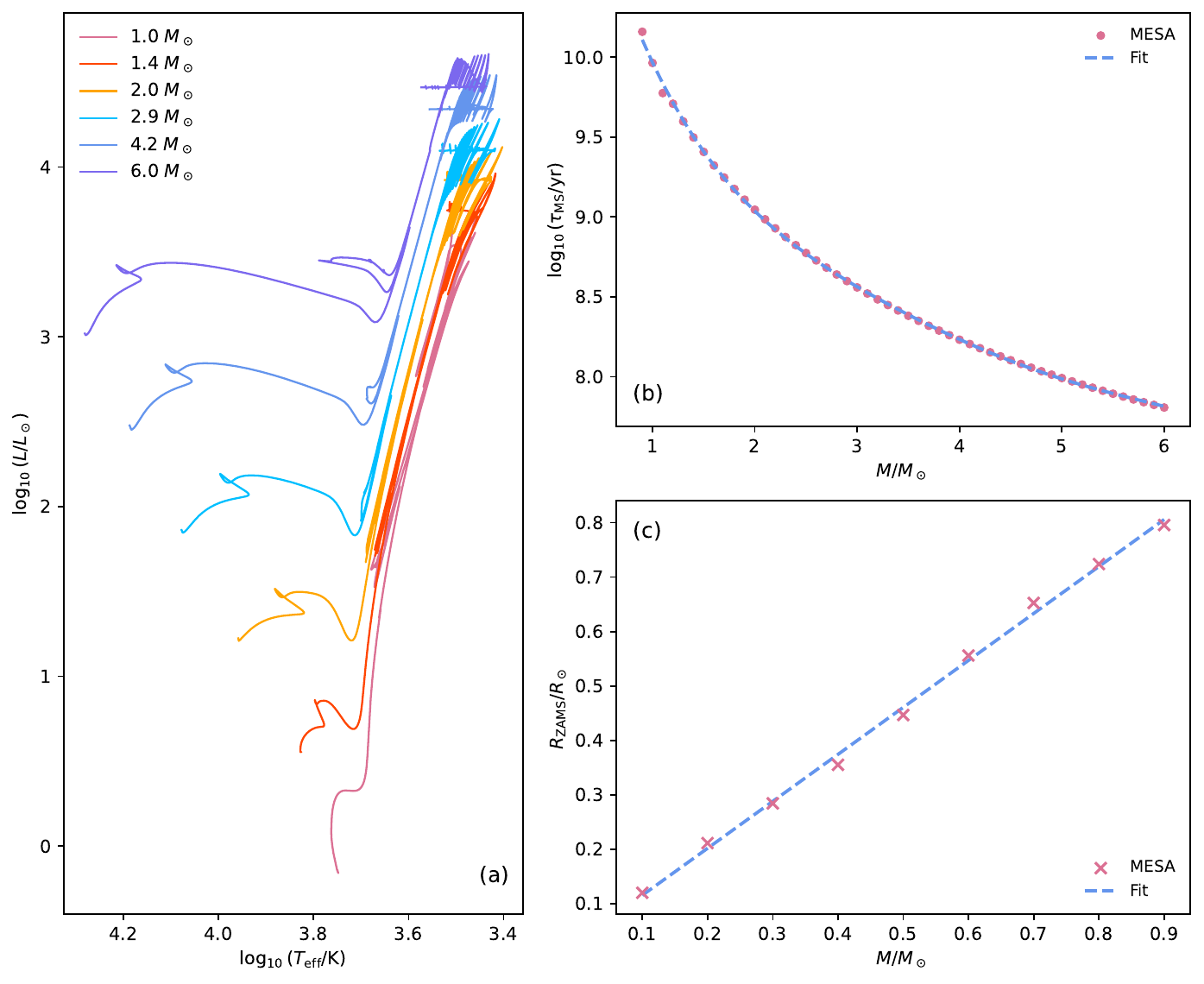}
\caption{Single-star models and fitting relations. (a) Hertzsprung--Russell diagram for six representative MESA tracks. (b) MS lifetimes for all 52 tracks and a fifth-order polynomial fit to the logarithmic lifetime as a function of initial mass. (c) ZAMS radii for stars with initial masses below $0.9\,M_\odot$ and a linear fit as a function of initial mass.}
\label{fig:MESA_grid}
\end{figure*}

We use version 26.4.1 of the stellar evolution code Modules for Experiments in Stellar Astrophysics (MESA) \citep{2011ApJS..192....3P,2013ApJS..208....4P,2015ApJS..220...15P,2018ApJS..234...34P,2019ApJS..243...10P,2023ApJS..265...15J} to construct 52 detailed single-star evolutionary tracks with initial masses from 0.9 to 6 $M_\odot$ in steps of 0.1 $M_\odot$. All models start on the pre-MS with an initial metallicity of $Z=0.02$ and no rotation. We evolve the $0.9\,M_\odot$ model through the end of the MS because its MS lifetime exceeds the current age of the Universe. The other 51 models are evolved through the end of the TP-AGB, which we define as the point at which the remaining envelope mass falls below $0.02\,M_\odot$. Their final core masses range from approximately 0.53 to 0.96 $M_\odot$. Figure~\ref{fig:MESA_grid}(a) shows six representative tracks.

We adopt OPAL opacity tables \citep{1996ApJ...464..943I}, including Type-II opacities for carbon- and oxygen-enhanced compositions, together with low-temperature opacity tables from \citet{2005ApJ...623..585F}, using the GS98 abundance mixture \citep{1998SSRv...85..161G}. Nuclear reaction rates follow the MESA defaults, primarily based on the JINA REACLIB database \citep{2010ApJS..189..240C}. We use the default time-dependent convection prescription with a baseline mixing-length parameter of $\alpha_{\rm MLT}=2.0$.

We adopt exponential diffusive overshooting, with parameters ($f,\,f_0$) = (0.016, 0.008) for convective cores and (0.0174, 0.0087) at the bottom boundaries of convective shells \citep{2016ApJ...823..102C}. For the pulse-driven convection zone during the TP-AGB phase, we adopt weaker overshooting with ($f,\,f_0$) = (0.008, 0.0008), motivated by \citet{2008IAUS..252..205H}. Stellar wind mass loss is described by the \citet{1975MSRSL...8..369R,1977A&A....61..217R} prescription with $\eta_{\mathrm{R}}$ = 0.5 before central helium depletion and the \citet{1995A&A...297..727B} prescription with $\eta_{\mathrm{B}}$ = 0.05 thereafter \citep{2014ApJ...782...17K}. To resolve the rapid structural variations and thin burning shells associated with thermal pulses, we tighten the temporal and spatial resolution after central helium depletion. For models with initial masses of at least 2.2 $M_\odot$, we additionally employ a modified abundance-gradient mesh prescription to improve the resolution and numerical convergence of TP-AGB evolution. Our MESA files and evolutionary tracks are
available on Zenodo doi: 10.5281/zenodo.23916438.

The late TP-AGB phase is particularly challenging to model because the hydrogen recombination instability and Fe-peak instability can prevent one-dimensional hydrostatic models from evolving through the thermal pulses \citep{2012A&A...542A...1L,2024MNRAS.531.4033R}. Following \citet{2024MNRAS.531.4033R}, we implement customized treatments to suppress these instabilities. Specifically, we damp the gravothermal energy term associated with hydrogen recombination in the cool outer envelope and enhance convective energy transport in regions where radiation pressure becomes dominant. Our tests confirm that these treatments have negligible effects on the stellar evolution prior to the onset of the instabilities, consistent with the findings of \citet{2024MNRAS.531.4033R}. These treatments are intended only to enable the models to continue evolving through the late TP-AGB phase under the adopted wind mass-loss prescriptions rather than accurately describing the actual dynamical evolution of the envelope. Their effects are further discussed in Section~\ref{sec:discussion}.

To incorporate these detailed stellar models into our BPS framework, we preprocess the evolutionary tracks and retain only the evolution following the ZAMS. The ZAMS is identified as the first point where nuclear burning supplies at least 95\% of the total stellar luminosity, and the stellar age is reset to zero at this point. For each evolutionary step, we extract the stellar age, mass, helium-core mass, radius, luminosity, effective temperature, and wind mass-loss rate. In addition, we calculate the envelope gravitational energy ($E_{\mathrm{grav}}$), internal energy ($E_{\mathrm{int}}$), convective-envelope mass ($M_{\mathrm{env,c}}$) and radial extent ($R_{\mathrm{env,c}}$), and total stellar moment of inertia ($I$), which are subsequently used to describe CEE, tidal interactions, and stellar rotation in the BPS model.

The envelope energies are calculated directly from the stellar structure after central helium depletion, by integrating from the helium-core boundary to the stellar surface:
\begin{equation}
E_{\rm grav}=-\int_{M_{\rm c}}^{M}\frac{Gm}{r(m)}\,dm,
\label{eq:egrav}
\end{equation}
\begin{equation}
E_{\rm int}=\int_{M_{\rm c}}^{M}u(m)\,dm,
\label{eq:eint}
\end{equation}
where $G$ is the gravitational constant, $M_{\mathrm{c}}$ is the helium-core mass, and $r(m)$ is the radius at mass coordinate $m$. The helium-core boundary is defined using the MESA core-boundary criterion corresponding to a hydrogen mass fraction of 0.1. The specific internal energy $u(m)$ is obtained from the MESA equation of state and includes contributions from the thermal energy of gas, radiation, ionization/recombination, molecular dissociation, and non-ideal effects.

The convective-envelope mass and radial extent are determined from the boundaries of the convective region in the MESA models. The total stellar moment of inertia is calculated assuming spherical symmetry:
\begin{equation}
I=\frac{2}{3}\int_{0}^{M}r^2(m)\,dm.
\label{eq:moment_inertia}
\end{equation}

We fit the MS lifetimes of all 52 tracks, defining the end of the MS as the point at which the central hydrogen mass fraction falls below $10^{-4}$. The fifth-order fit shown in Figure~\ref{fig:MESA_grid}(b) has a coefficient of determination of $R^2=0.9996$ and is given by
\begin{equation}
\begin{aligned}
\log_{10}&[\tau_{\rm MS}(M)/\mathrm{yr}]= \\
& 12.1162618 - 3.16439614M \\
& + 1.27791734M^2 - 0.296499197M^3 \\
& + 0.0350925661M^4 - 0.00164054461M^5.
\end{aligned}
\label{eq:tau_MS}
\end{equation}

For stars with initial masses below $0.9\,M_\odot$, we use a linear fit for their ZAMS radii (Figure~\ref{fig:MESA_grid}(c)) and treat them as unevolved after the ZAMS. The fit is
\begin{equation}
R_{\rm ZAMS}(M)/R_\odot = 0.0292087 + 0.86294848M,
\label{eq:R_ZAMS}
\end{equation}
with $R^2=0.9972$. In Equations~\eqref{eq:tau_MS} and \eqref{eq:R_ZAMS}, $M$ is expressed in solar masses. We use these fits in the binary evolution calculations described in Section~\ref{sec:bps}, including companion rejuvenation and CEE.

\section{Binary Population Synthesis Model} \label{sec:bps}

\subsection{Overview} \label{sec:overview}

Our BPS framework combines the precomputed MESA donor tracks (Section~\ref{sec:stellar_grid}) with parametrized binary-interaction prescriptions (Section~\ref{sec:binary_interaction}). The primary follows its detailed track while the companion and orbit are updated simultaneously. Preserving the stellar-structure information and temporal resolution while substantially reducing computational cost allows us to resolve thermal pulses in millions of binaries. The open-source framework is freely available for use and further modification.\footnote{The source code is available in the \href{https://github.com/lonelylight1945/Thermal-Pulse-Resolved-Binary-Population-Synthesis}{project's GitHub repository}. The version used in this work has been archived on Zenodo (10.5281/zenodo.23916438).}

We generate $5\times10^6$ binaries through Monte Carlo sampling of the distributions in Table~\ref{tab:initial_distributions}, following \citet[Section~2.1]{2026ApJ..1007...15S} but restricting the initial primary mass to $1$--$6\,M_\odot$. Each sampled mass is assigned to the nearest grid point, $M'_{1\rm i}$, and the initial companion mass is $M_{2\rm i}=q_{\rm i}M'_{1\rm i}$, where $q_{\rm i}=M_{2\rm i}/M_{1\rm i}$. The primary masses are thus discretized while their relative numbers still follow the adopted IMF. The birth time $t_{\rm born}$ is measured from the formation of the Galaxy, and the maximum evolutionary time is $t_{\rm max}=13\,\mathrm{Gyr}-t_{\rm born}$, corresponding to a constant star-formation rate.

\begin{deluxetable}{ccc}[ht!]
\tabletypesize{\scriptsize}
\tablecaption{Initial binary parameter distributions.\label{tab:initial_distributions}}
\tablehead{
\colhead{Parameter} & \colhead{Range} & \colhead{Distribution}
}
\startdata
$M_{1\rm i}$ & $[1,6]\,M_\odot$ & Miller--Scalo IMF \\
$q_{\rm i}$ & $(0,1]$ & Flat \\
$a_{\rm i}$ & $(0,5.75\times10^6]\,R_\odot$ & Broken power law \\
$t_{\rm born}$ & $[0,13]$ Gyr & Flat \\
\enddata
\tablecomments{The distributions of $M_{\rm 1i}$ and $a$ are from \citet{1979ApJS...41..513M} and \citet{2020RAA....20..161H}, respectively. The companion mass is restricted to $M_{2\rm i}\geq0.1\,M_\odot$.}
\end{deluxetable}

At each step along the primary's track, we update the companion mass and evolutionary state, primary spin angular momentum, and orbital separation, accounting for wind mass loss, wind accretion, and tides. Orbits are initially circular and remain so throughout detached evolution.

Evolution stops at RLOF, the maximum evolutionary time, the companion leaving the MS, or the end of the donor track without RLOF. We consider only TP-AGB-initiated CEE. In the standard model, all RLOF during this phase immediately initiates CEE, whose outcome follows the energy formalism in Section~\ref{sec:ce_evolution}. We retain systems that successfully eject the envelope and contain an MS companion at the present epoch, without following their subsequent post-CE orbital evolution.

\subsection{Binary Interaction} \label{sec:binary_interaction}

\subsubsection{Mass Evolution} \label{sec:mass_evolution}

The mass evolution of the primary is directly determined by its precomputed MESA track. During the detached phase, mass loss from the primary occurs through stellar winds, part of which may be accreted by the companion. Given the strong winds during the TP-AGB phase, we account for wind accretion and the associated rejuvenation of the MS companion.

For circular orbits, the companion's wind accretion rate is calculated using the Bondi--Hoyle prescription \citep{1944MNRAS.104..273B,2002MNRAS.329..897H}:
\begin{equation}
\dot{M}_{\rm a}=-\frac{\alpha_{\rm w}}{2a^2}\left(\frac{GM_{\rm a}}{v_{\rm w}^2}\right)^2\left(1+\frac{v_{\rm orb}^2}{v_{\rm w}^2}\right)^{-3/2}\dot{M}_{\rm w},
\label{eq:wind_accretion}
\end{equation}
where $M_{\mathrm{a}}$ is the accretor mass, $a$ is the orbital separation, and ${\dot{M}}_{\mathrm{w}}$ denotes the donor wind mass-loss rate. We adopt ${\alpha}_{\mathrm{w}}=1.5$, following \citet{1988A&A...205..155B}. The squared wind and orbital velocities are given by:
\begin{equation}
v_{\rm w}^2=2\beta_{\rm w}\frac{GM_{\rm d}}{R_{\rm d}},
\label{eq:wind_velocity}
\end{equation}
\begin{equation}
v_{\rm orb}^2=\frac{G(M_{\rm d}+M_{\rm a})}{a},
\label{eq:orbital_velocity}
\end{equation}
where ${M}_{\mathrm{d}}$ and ${R}_{\mathrm{d}}$ are the donor mass and radius, respectively. We adopt ${\beta}_{\mathrm{w}}=1/8$, following \citet{2002MNRAS.329..897H}. The donor quantities are taken directly from the detailed stellar evolution grid. The accretion rate is limited to 80\% of the donor wind mass-loss rate as a numerical safeguard.

As the companion gains mass, its evolutionary state must be adjusted to account for rejuvenation. We define the MS evolutionary phase as:
\begin{equation}
f=\frac{t_{\rm eff}}{\tau_{\rm MS}(M)},
\label{eq:ms_phase}
\end{equation}
where ${t}_{\mathrm{eff}}$ is the effective stellar age and ${\tau}_{\mathrm{MS}}(M)$ is the MS lifetime at the current stellar mass, calculated from Equation~\ref{eq:tau_MS}. Thus, $f=0$ corresponds to the ZAMS and $f=1$ to the end of the MS.

For MS stars with radiative cores, approximately in the range $0.9$--$1.15\,M_\odot$ in our models, we follow the rejuvenation treatment of \citet{1997MNRAS.291..732T} and \citet{2000MNRAS.315..543H,2002MNRAS.329..897H}. After accretion, the effective age is rescaled according to the change in MS lifetime: $t_{\rm eff}'=[\tau_{\rm MS}(M')/\tau_{\rm MS}(M)]t_{\rm eff}$, where $M$ and $M'$ denote the companion masses before and after accretion. Consequently, $f'=f$; the MS phase remains unchanged, although the remaining MS lifetime decreases.

For stars with convective cores, which appear above approximately $1.15\,M_{\odot}$ in our models, mass accretion increases the convective-core mass and allows fresh hydrogen to be mixed into the core. Following \citet{1997MNRAS.291..732T} and \citet{2007MNRAS.376...61D}, we adopt $f'=\frac{M}{M'}f$. Since $M'>M$, accretion reduces the evolutionary phase, making the companion effectively younger.

For companions below $0.9\,M_{\odot}$, whose MS lifetimes greatly exceed the age of the Universe, we approximate their evolutionary states as remaining at the ZAMS until their masses exceed this threshold.

\subsubsection{Spin Evolution} \label{sec:spin_evolution}

We assign the primary an initial rotation velocity as a function of its ZAMS mass, following \citet{2000MNRAS.315..543H} and \citet{1992adps.book.....L}:
\begin{equation}
v_{\rm rot,i}(M)=\frac{330M_{1\rm i}^{3.3}}{15+M_{1\rm i}^{3.45}}\ {\rm km\,s^{-1}},
\label{eq:initial_rotation}
\end{equation}
where $M_{1\rm i}$ is expressed in solar masses. The initial angular velocity and spin angular momentum are ${\Omega}_{\mathrm{spin,i}}={v}_{\mathrm{rot,i}}/{R}_{\mathrm{ZAMS}}$ and ${J}_{\mathrm{spin,i}}=I{\Omega}_{\mathrm{spin,i}}$, where the moment of inertia is obtained from the detailed stellar evolution grid.

During detached binary evolution, we consider changes in the primary's spin angular momentum caused by stellar winds and tidal interactions:
\begin{equation}
\dot{J}_{\rm spin}=\dot{J}_{\rm spin,w}+\dot{J}_{\rm spin,t}.
\label{eq:spin_evolution}
\end{equation}

Stellar wind mass loss removes spin angular momentum from the donor at the rate \citep{2000MNRAS.315..543H}:
\begin{equation}
\dot{J}_{\rm spin,w}=\frac{2}{3}\dot{M}_{\rm w}R^2\Omega_{\rm spin},
\label{eq:wind_spin}
\end{equation}
where the negative sign of ${\dot{M}}_{\mathrm{w}}$ represents angular momentum loss.

Tidal interactions transfer angular momentum between the donor's spin and the binary orbit. For circular orbits, the tidal torque on the donor is calculated as \citep{1981A&A....99..126H,2002MNRAS.329..897H}:
\begin{equation}
\begin{aligned}
\dot{J}_{\rm spin,t} &= I\frac{d\Omega_{\rm spin,t}}{dt} \\
&=3\left(\frac{k}{T}\right)\frac{q^2}{r_{\rm g}^2}\left(\frac{R}{a}\right)^6(\Omega_{\rm orb}-\Omega_{\rm spin})I,
\label{eq:tidal_torque}
\end{aligned}
\end{equation}
where $q=M_{\rm a}/M_{\rm d}$, ${r}_{\mathrm{g}}^{2}=I/({M}_{\mathrm{d}}{R}^{2})$ is the squared gyration radius, and $k/T$ is a combined quantity characterizing the efficiency of tidal dissipation.

For stars with substantial convective envelopes, which in our models mainly include donors on the RGB and AGB, as well as those in the core-helium-burning phase with initial masses of $\lesssim 4.5\,M_\odot$, we adopt equilibrium tides with convective damping \citep{1981A&A....99..126H,2002MNRAS.329..897H}:
\begin{equation}
\left(\frac{k}{T}\right)=\frac{2}{21}\frac{f_{\rm conv}}{\tau_{\rm conv}}\frac{M_{\rm env,c}}{M}\ {\rm yr}^{-1},
\label{eq:convective_tide}
\end{equation}
where ${M}_{\mathrm{env,c}}$ is the convective-envelope mass, ${\tau}_{\mathrm{conv}}$ is the convective turnover timescale, and ${f}_{\mathrm{conv}}$ is the tidal dissipation reduction factor. Following \citet{2002MNRAS.329..897H},
\begin{equation}
\begin{aligned}
\tau_{\rm conv}&=0.4311\\
&\times\left[\frac{M_{\rm env,c}R_{\rm env,c}(R-0.5R_{\rm env,c})}{3L}\right]^{1/3},
\label{eq:convective_timescale}
\end{aligned}
\end{equation}
\begin{equation}
P_{\rm tide}=\frac{2\pi}{\lvert\Omega_{\rm orb}-\Omega_{\rm spin}\rvert},
\label{eq:tidal_period}
\end{equation}
\begin{equation}
f_{\rm conv}=\min\left[1,\left(\frac{P_{\rm tide}}{2\tau_{\rm conv}}\right)^2\right].
\label{eq:convective_reduction}
\end{equation}
Here, the stellar mass, radius, luminosity ($L$), and convective-envelope properties are obtained from the detailed stellar evolution grid (see Section~\ref{sec:stellar_grid}). Solar units are used in the convective turnover timescale prescription, yielding ${\tau}_{\mathrm{conv}}$ in years.

For stars with radiative envelopes (in our current implementation, this applies to the MS phase and to core-helium-burning models with initial masses of $\gtrsim4.5\,M_\odot$), we adopt the dynamical-tide prescription with radiative damping \citep{1975A&A....41..329Z,1977A&A....57..383Z,2002MNRAS.329..897H}.\footnote{Equation (42) of \citet{2002MNRAS.329..897H} contains a typographical error; the corrected form is used here.}
\begin{equation}
\left(\frac{k}{T}\right)=\sqrt{\frac{GMR^2}{a^5}}(1+q)^{5/6}E_2,
\label{eq:radiative_tide}
\end{equation}
where the masses and radii are expressed in solar units, and ${E}_{2}=1.592\times {10}^{-9}{M}^{2.84}$.

We neglect the spin evolution of the companion. For the wide TP-AGB binaries considered here, the spin angular momentum of a low-mass MS companion is typically estimated to be of order ${10}^{-4}$ or less of the total angular momentum, so neglecting its contribution is expected to have little effect on the orbital evolution. The primary's spin is also limited to its critical rotation rate, with any excess spin angular momentum assumed to be removed through mass loss.

\subsubsection{Orbital Evolution} \label{sec:orbital_evolution}

The binary orbit evolves in response to stellar wind mass loss, wind accretion, and tidal interactions. The evolution of orbital angular momentum is described by

\begin{equation}
\dot{J}_{\rm orb}=\dot{J}_{\rm orb,w}+\dot{J}_{\rm orb,a}+\dot{J}_{\rm orb,t}.
\label{eq:orbital_evolution}
\end{equation}

When a star loses mass through stellar wind, it not only carries away its own rotational angular momentum but also removes some of the orbital angular momentum. Following \citet{2002MNRAS.329..897H}, the orbital angular momentum carried away by stellar winds is
\begin{equation}
\dot{J}_{\rm orb,w}=\dot{M}_{\rm w}\left(\frac{M_{\rm a}}{M_{\rm d}+M_{\rm a}}\right)^2a^2\Omega_{\rm orb},
\label{eq:wind_orbital_am}
\end{equation}
and the additional contribution associated with wind accretion is
\begin{equation}
\dot{J}_{\rm orb,a}=-\dot{M}_{\rm a}\frac{M_{\rm a}M_{\rm d}}{(M_{\rm d}+M_{\rm a})^2}a^2\Omega_{\rm orb}.
\label{eq:accretion_orbital_am}
\end{equation}

Tides only exchange angular momentum between spin and orbit, so conservation requires
\begin{equation}
\dot{J}_{\rm orb,t}=-\dot{J}_{\rm spin,t}.
\label{eq:tidal_orbital_am}
\end{equation}

The orbital separation and Roche-lobe radius are updated from the evolving orbital angular momentum and component masses throughout the donor's evolution, including individual thermal-pulse cycles.

\subsubsection{Common-envelope Evolution} \label{sec:ce_evolution}

We calculate the outcome of common-envelope evolution using the energy formalism \citep{1984ApJ...277..355W,1988ApJ...329..764L}. In this prescription, orbital energy released during the inspiral is used to overcome the binding energy of the donor's envelope. The post-CE orbital separation is determined by
\begin{equation}
\alpha_{\rm CE}\left(\frac{GM_{\rm d}M_{\rm a}}{2a_{\rm i}}-\frac{GM_{\rm c}M_{\rm a}}{2a_{\rm f}}\right)=E_{\rm bind},
\label{eq:ce_energy}
\end{equation}
where ${M}_{\mathrm{d}}$, ${M}_{\mathrm{c}}$, and $M_{\mathrm{a}}$ are the donor mass, core mass, and the companion mass at the onset of CEE, respectively, $a_{\mathrm{i}}$ and $a_{\mathrm{f}}$ are the pre- and post-CE orbital separations. The parameter ${\alpha}_{\mathrm{CE}}$ represents the efficiency with which the released orbital energy contributes to envelope ejection. The envelope binding energy is calculated as
\begin{equation}
E_{\rm bind}=E_{\rm grav}+\alpha_{\rm th}E_{\rm int},
\label{eq:binding_energy}
\end{equation}
where ${\alpha}_{\mathrm{th}}$ specifies the fraction of internal energy contributing to envelope ejection.\footnote{Unlike some BPS studies \citep[e.g.,][]{2022ApJS..258...34R,2026ApJ..1007...15S}, where $\alpha_{\rm th}$ serves as a linear interpolation weight between $\lambda_{\rm g}$ and $\lambda_{\rm b}$, our definition directly scales $E_{\rm int}$. The former does not correspond to the actually included fraction of internal energy, because $E_{\rm bind}\propto 1/\lambda$.} Both ${E}_{\mathrm{grav}}$ and ${E}_{\mathrm{int}}$ are obtained directly from the detailed stellar structure at the onset of CEE, as described in Section~\ref{sec:stellar_grid}. During the late TP-AGB phase, $E_{\rm bind}$ may become positive, depending on the adopted value of ${\alpha}_{\mathrm{th}}$. We exclude such configurations because the $\alpha$-formalism is no longer applicable \citep{2024PASP..136h4202Y}. 

For systems with negative $E_{\rm bind}$, we calculate the companion's Roche-lobe radius, $R_{\rm L2}$, using the post-CE separation $a_{\rm f}$ obtained from Equation~\ref{eq:ce_energy}. Systems with $R_{\rm L2}\leq R_2$ are classified as mergers; otherwise, they survive as detached PCEBs. Here, $R_2$ is the companion radius at the onset of CEE, interpolated from the detailed stellar evolution grid described in Section~\ref{sec:stellar_grid} using its current mass and MS evolutionary phase $f$ for $M_2\geq0.9\,M_\odot$, or taken as the ZAMS radius calculated from Equation~\ref{eq:R_ZAMS} using its current mass for $M_2<0.9\,M_\odot$.

Unlike calculations based on other analytic fitting formulae for $E_{\rm bind}$ \citep[e.g.,][]{2010ApJ...716..114X,2010ApJ...722.1985X,2011ApJ...743...49L,2016RAA....16..126W}, our approach uses the real instantaneous donor structure at the onset of CEE. This allows the changes in $E_{\rm bind}$ associated with individual thermal pulses to be incorporated directly and self-consistently into the predicted post-CE orbital separations. We discuss the significance of this point for TP-AGB CEE in Section~\ref{sec:pulse_effects}.

\section{Effects of Thermal Pulses on Binary Evolution} \label{sec:pulse_effects}

\begin{figure*}[ht!]
\centering
\includegraphics[width=\textwidth]{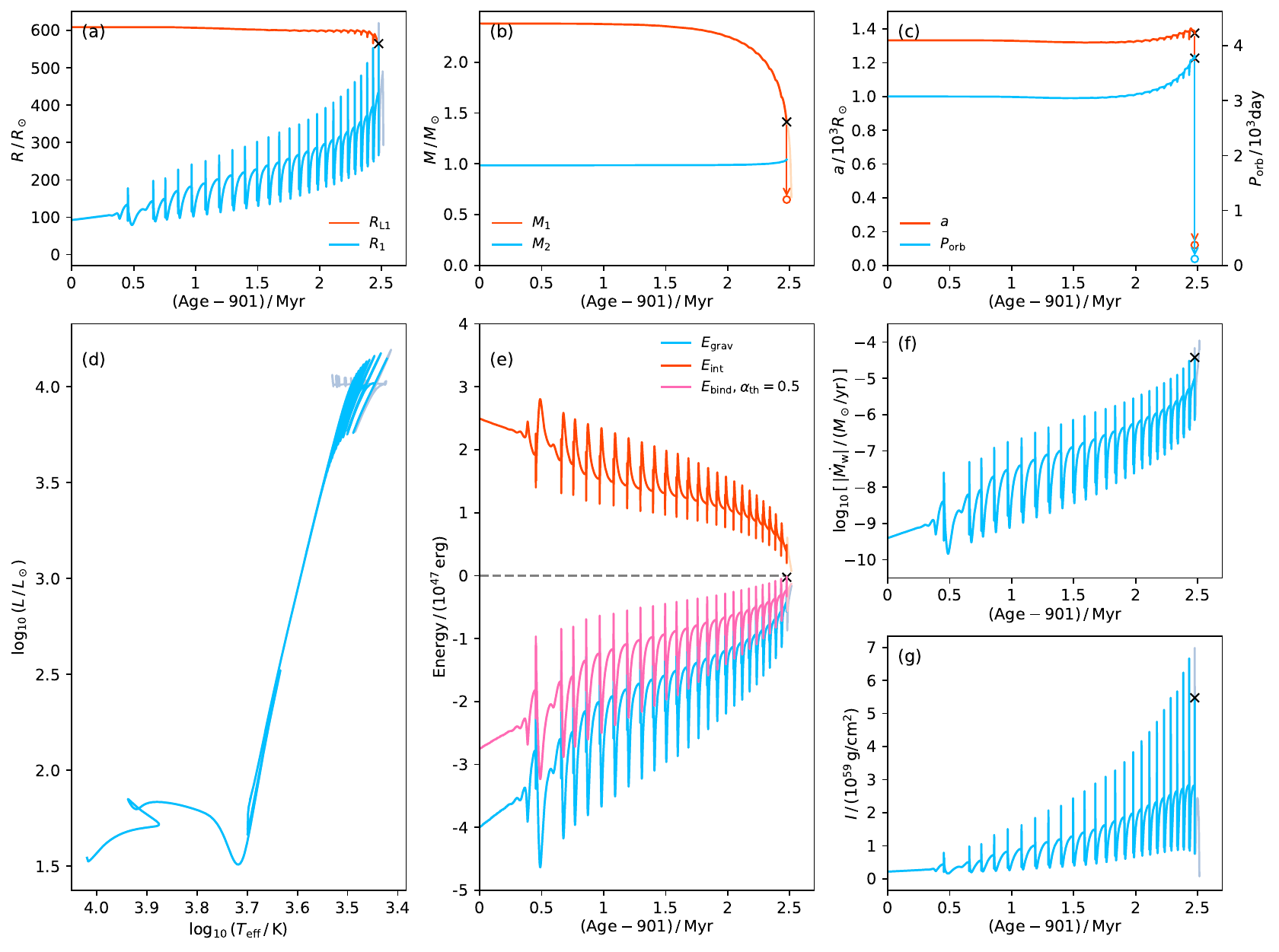}
\caption{Evolutionary history of a representative binary with initial parameters ${M}_{1\rm i}=2.4\,M_\odot$, ${M}_{2\rm i}=0.9852\,M_\odot$, and ${a}_{\mathrm{i}}=1325.66\,R_\odot$. The panels show (a) the donor radius and its Roche-lobe radius, (b) the donor and companion masses, (c) the orbital separation and orbital period, (d) the Hertzsprung--Russell diagram, (e) the envelope gravitational energy, internal energy, and binding energy, (f) the donor wind mass-loss rate, and (g) the donor's total moment of inertia. For clarity, only the TP-AGB phase is shown in panels other than (d). Black crosses mark the onset of CEE, while open circles indicate the post-CE configuration. Dark curves represent the pre-RLOF binary evolution, whereas lighter curves show the continuation of the corresponding single-star donor track. For ${\alpha}_{\mathrm{CE}}=0.3$ and ${\alpha}_{\mathrm{th}}=0.5$, the system forms a PCEB with ${M}_{\mathrm{WD}}=0.648\,M_\odot$, ${M}_{2}=1.042\,M_\odot$, and ${P}_{\mathrm{orb}}=118.35$ days.}
\label{fig:evolution}
\end{figure*}

We begin by examining how individual thermal pulses affect the onset of RLOF and the outcome of CEE. To illustrate these effects, we follow a representative binary from the ZAMS to PCEB formation. The system has initial component masses ${M}_{1\rm i}=2.4\,M_\odot$ and ${M}_{2\rm i}=0.9852\,M_\odot$ and an initial orbital separation of ${a}_{\mathrm{i}}=1325.66\,R_\odot\,(6.16\,\mathrm{AU})$. Figure~\ref{fig:evolution} shows the evolution with time of important properties of the donor's radius, mass, wind mass-loss rate, moment of inertia, envelope energies, and the orbital separation.

Before the TP-AGB, the primary evolves through the MS, subgiant, RGB, and core-helium-burning phases with relatively small changes in the stellar masses and orbit. During the TP-AGB, helium accumulated by hydrogen-shell burning repeatedly ignites in thermally unstable shell flashes, driving structural changes and envelope expansion (Figure~\ref{fig:evolution}(a)). The increasing radius excursions permit RLOF in otherwise detached wide binaries. Here, after 25 pulses, the donor fills its Roche lobe on the rising branch of the final pulse and enters CEE. Smoothing out these radius variations could incorrectly classify systems that interact only during individual pulses as detached.

The donor's wind mass-loss rate and moment of inertia also vary strongly between pulses and increase substantially through the TP-AGB (Figures~\ref{fig:evolution}(f) and \ref{fig:evolution}(g)). Wind mass loss generally expands the orbit, partly offset by wind accretion, while tides exchange angular momentum between spin and orbit according to their relative angular velocities (Section~\ref{sec:orbital_evolution}). During the early TP-AGB, winds are weak; increasing radius and moment of inertia tend to spin down the donor, allowing tides to transfer orbital angular momentum to its spin and slightly contract the orbit. Later, the wind mass-loss rate increases by approximately three orders of magnitude, and wind-driven expansion dominates tidal shrinkage. The separation therefore first decreases slightly and then increases toward the end of the TP-AGB (Figure~\ref{fig:evolution}(c)), while wind accretion modestly increases the companion mass (Figure~\ref{fig:evolution}(b)).

An additional effect becomes apparent when the donor approaches Roche-lobe filling. As shown in Figures~\ref{fig:evolution}(a) and \ref{fig:evolution}(c), the orbital separation and the donor's Roche-lobe radius also vary during thermal pulses, with their minima occurring near the maxima of the stellar radius. This behavior arises because the rapid envelope expansion increases the stellar moment of inertia and tends to reduce its spin angular velocity, while tidal interactions transfer angular momentum from the orbit to the donor. Unlike the gradual orbital contraction occurring during the early TP-AGB, this process takes place over the much shorter timescale of an individual thermal pulse, typically a few hundred years. Since the tidal torque depends strongly on the ratio $R/a$ (Section~\ref{sec:spin_evolution}), appreciable orbital shrinkage over such a short interval occurs primarily when the donor is already close to filling its Roche lobe. This tidally driven approach to interaction is similar to the tidal capture of companions by giant stars discussed by \citet{2016MNRAS.463.1040M}; comparable behaviour is also reported in \citet{2024A&A...686A.226B}. Consequently, thermal-pulse expansion and tidal orbital shrinkage can act together to promote RLOF, extending the range of initial binary configurations that may undergo interaction.

Both gravitational and internal energies respond to individual pulses, while the binding-energy magnitude decreases overall as the donor loses its envelope (Figure~\ref{fig:evolution}(e)). The envelope can become substantially less bound during pulse-driven expansion, precisely when RLOF is most likely to occur. Here, the preceding evolution and final pulse have already greatly reduced the binding-energy magnitude at RLOF, so relatively modest orbital contraction suffices to eject the envelope and leave a PCEB with ${P}_{\mathrm{orb}}=118.35$ days.

Conventional fitting formulae for $E_{\rm bind}$ \citep[e.g.,][]{2010ApJ...716..114X,2010ApJ...722.1985X,2011ApJ...743...49L,2016RAA....16..126W} do not explicitly resolve these rapid variations across individual thermal pulses. They therefore cannot accurately capture the $E_{\rm bind}$ at the onset of RLOF, particularly when the interaction is triggered by a thermal pulse. Since MT preferentially occurs during thermal pulses \citep{2024A&A...687A..12B,2024PASP..136h4202Y}, only the binding energy during thermal pulses is effective. Adopting time-averaged binding energies or extrapolating prescriptions from the early AGB phase may overestimate the magnitude of the binding energy at the onset of CEE, potentially making the formation of long-period PCEBs more difficult in rapid BPS calculations \citep[e.g.,][]{2026ApJ..1007...15S}. More importantly, simply introducing a more accurate binding-energy prescription into a rapid BPS code would not capture the full effect of thermal pulses, because the onset of interaction also depends on the coupled evolution of stellar radius, wind mass loss, stellar spin, tides, and orbital separation. A pulse-resolved treatment of these processes is therefore essential for investigating their combined influence on the resulting PCEB population.

\section{Thermal-pulse-resolved BPS Results} \label{sec:bps_results}

\subsection{Properties of the Synthetic Population} \label{sec:population}

\begin{figure*}[ht!]
\centering
\includegraphics[width=0.9\textwidth]{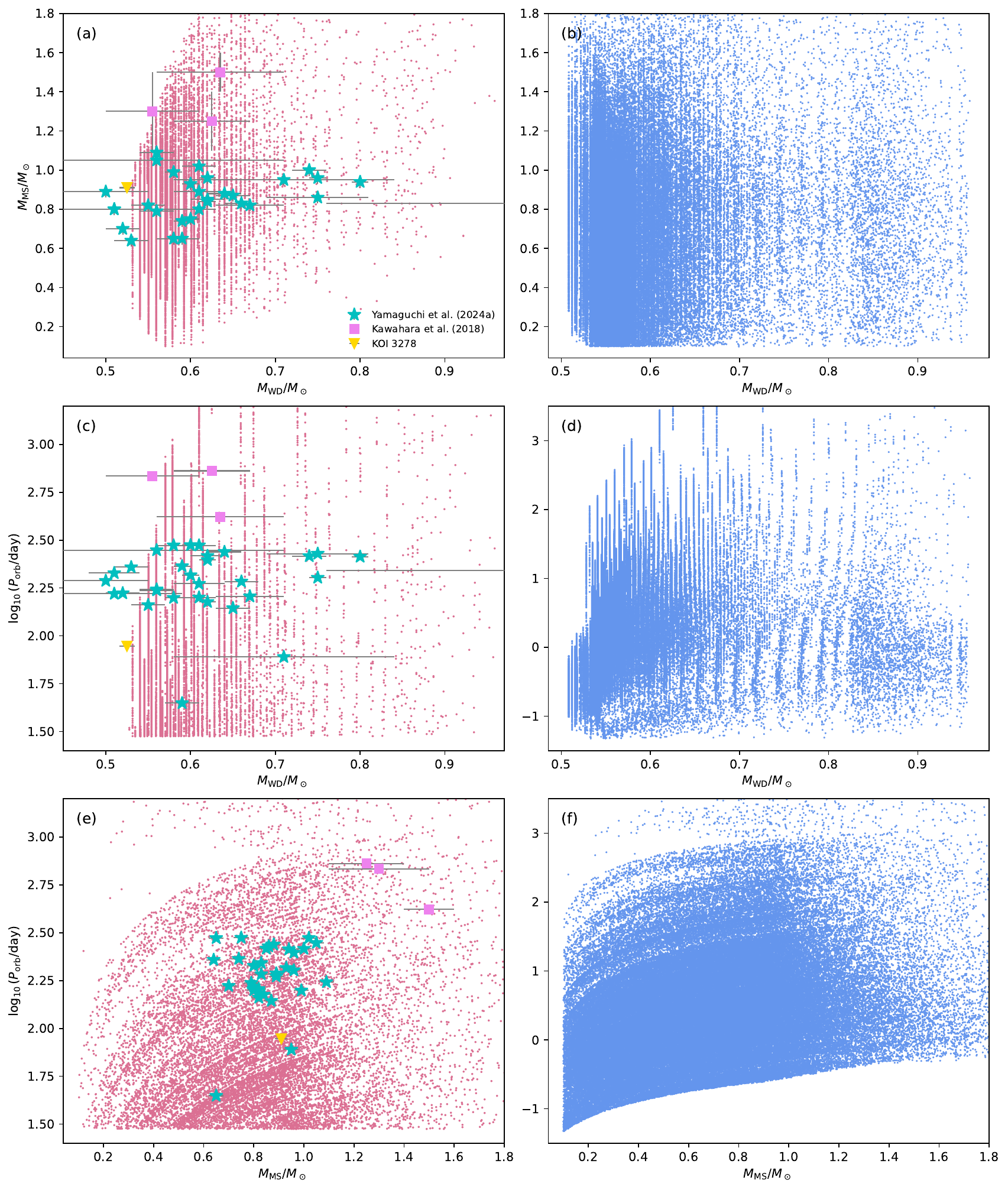}
\caption{Distributions of the synthetic PCEB population obtained with ${\alpha}_{\mathrm{CE}}=0.3$ and ${\alpha}_{\mathrm{th}}=0.5$. Blue points show the full synthetic population, while pink points represent long-period systems with ${P}_{\mathrm{orb}}>30$ days. Observed long-period WD+MS systems are taken from \citet{2018AJ....155..144K} and \citet{2024PASP..136h4202Y}. For KOI-3278, discovered by \citet{2014Sci...344..275K}, we adopt the updated parameters from \citet{2019ApJ...880...33Y}.}
\label{fig:population}
\end{figure*}

Figure~\ref{fig:population} presents the synthetic PCEB population obtained with our standard parameters, ${\alpha}_{\mathrm{CE}}=0.3$ and ${\alpha}_{\mathrm{th}}=0.5$. The resulting orbital periods can span a broad range, from less than one day to more than $10^{3}$ days, covering both short- and long-period systems. Adopting ${P}_{\mathrm{orb}}>30$ days as the definition of long-period PCEBs, we find that these systems account for approximately 11.5\% of the synthetic population. This fraction may increase if subsequent post-CE evolution is included, since angular momentum loss through magnetic braking can drive some short-period PCEBs into the cataclysmic variables phase \citep{2014A&A...568A..68Z,2024A&A...682A..33B,2024PASP..136l4201B,2026ApJ..1007...15S}. The predicted long-period population also occupies a broad range of WD masses, companion masses, and orbital periods, overlapping with the observed WD+MS systems shown in Figure~\ref{fig:population}. The vertical stripes visible in the WD mass distributions arise from the discretization of the initial primary masses onto the MESA grid points (Sections~\ref{sec:stellar_grid} and \ref{sec:overview}), rather than from physical features of the population.

\begin{figure}[ht!]
\centering
\includegraphics[width=\columnwidth,height=0.72\textheight,keepaspectratio]{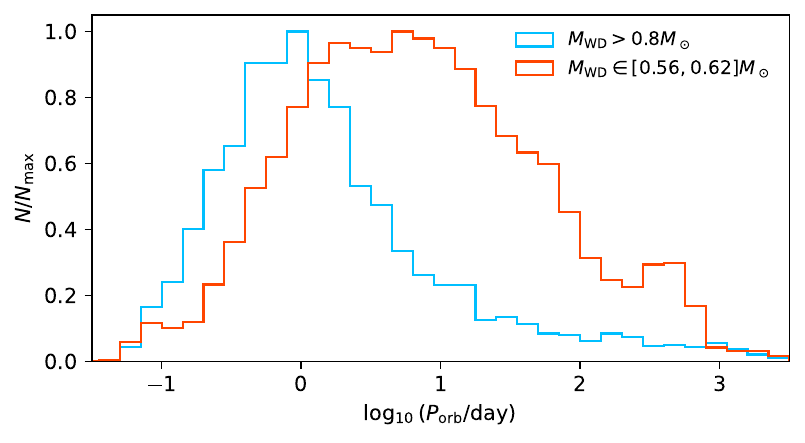}
\caption{Normalized orbital-period distributions of synthetic PCEBs with ${M}_{\mathrm{WD}}>0.8\,M_\odot$ (blue) and $0.56\leq {M}_{\mathrm{WD}}/M_\odot\leq 0.62$ (orangered), selected from the population shown in Figure~\ref{fig:population}(d). Each distribution is normalized to its maximum bin count to facilitate comparison of their shapes.}
\label{fig:period_distribution}
\end{figure}

A prominent feature of the long-period population is the strong concentration of WD masses around $0.6\,M_\odot$, with relatively few systems containing more massive WDs (Figures~\ref{fig:population}(a) and \ref{fig:population}(c)). In contrast, the overall synthetic population extends to considerably higher WD masses, although these systems preferentially occupy shorter orbital periods (Figure~\ref{fig:population}(d)). To illustrate this difference, Figure~\ref{fig:period_distribution} compares the orbital-period distributions of systems with $0.56\leq {M}_{\mathrm{WD}}/M_\odot\leq 0.62$ and ${M}_{\mathrm{WD}}>0.8\,M_\odot$. The former has a broad period distribution extending to hundreds of days, whereas the latter are concentrated mainly at shorter periods, with a peak around one day.

The relative scarcity of massive WDs in long-period PCEBs can be understood through the combined effects of envelope binding energy and companion evolution. As discussed in Section~\ref{sec:pulse_effects}, forming a long-period PCEB generally requires the donor to reach a highly evolved TP-AGB stage, where extensive wind mass loss has considerably reduced the envelope binding energy. However, the envelopes of more massive WD progenitors remain more strongly bound than those of lower-mass progenitors, even during the late TP-AGB. A sufficiently massive companion can provide a larger amount of orbital energy during CEE, allowing such systems to retain wider post-CE orbits. This explains the tendency for long-period systems containing massive WDs to have relatively massive companions (Figure~\ref{fig:population}(a)), while no comparable correlation is seen in the full PCEB population (Figure~\ref{fig:population}(b)).

Such companions, however, have shorter MS lifetimes and can survive as present-day WD+MS binaries only if the systems formed sufficiently recently. Their short lifetimes therefore limit the accumulation of long-period systems containing massive WDs over the star-formation history. By comparison, long-period systems hosting WDs near $0.6\,M_\odot$ have a median initial companion mass of approximately $0.81\,M_\odot$, allowing them to remain in the WD+MS phase over a much wider range of formation times. For more massive WD progenitors with low-mass companions, the greater envelope binding energy instead leads to stronger orbital shrinkage during CEE and hence shorter final periods. Together with the lower initial occurrence of massive WD progenitors, these effects strongly suppress the formation of present-day long-period WD+MS binaries containing massive WDs.

Observationally, a similar deficit has been reported by \citet{2024ApJ...970L..11H}, who found that the fraction of massive WDs in Gaia astrometric WD+MS binaries is substantially lower than that among field WDs, with the distribution concentrated toward approximately $0.6\,M_\odot$. This resembles the mass distribution predicted by our model, suggesting that late TP-AGB CEE may contribute to the observed scarcity of massive WDs in relatively wide WD+MS binaries.

\subsection{Dependence of Long-period PCEB Formation on CE Parameters} \label{sec:ce_parameters}

\begin{figure}[ht!]
\centering
\includegraphics[width=\columnwidth, keepaspectratio]{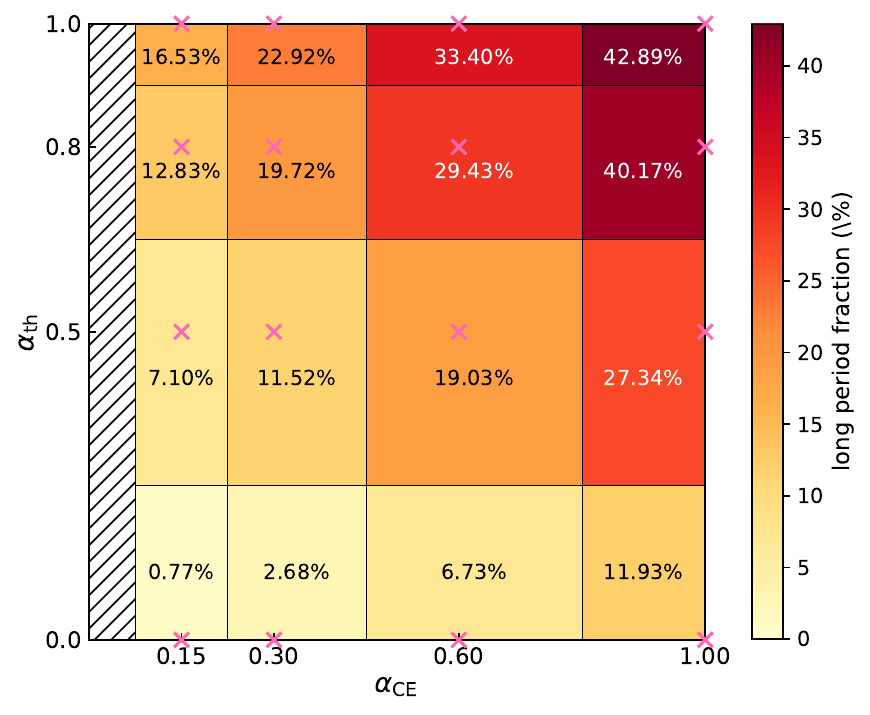}
\caption{Fraction of long-period systems (${P}_{\mathrm{orb}}>30\,\mathrm{days}$) among all synthetic PCEBs for different combinations of ${\alpha}_{\mathrm{CE}}$ and ${\alpha}_{\mathrm{th}}$. The percentage in each cell is calculated from the corresponding BPS model and assigned to the nearest-neighbor region in parameter space. Pink crosses indicate the sampled parameter combinations, while hatched regions lie outside the sampled range.}
\label{fig:ce_parameters}
\end{figure}

To investigate whether the formation of long-period PCEBs is general or depends on the specific CE parameters, we perform BPS calculations for 16 combinations of ${\alpha}_{\mathrm{CE}}$ and ${\alpha}_{\mathrm{th}}$. Figure~\ref{fig:ce_parameters} shows the fraction of systems with ${P}_{\mathrm{orb}}>30$ days among all synthetic PCEBs for each parameter combination. Long-period systems are produced throughout the explored parameter space, although their relative abundance varies substantially. The fraction is lowest for inefficient CE ejection with no contribution from internal energy, reaching only 0.77\%. It increases with both parameters and reaches 42.89\% at ${\alpha}_{\mathrm{CE}}=1$ and ${\alpha}_{\mathrm{th}}=1$.

\begin{figure}[ht!]
\centering
\includegraphics[width=\columnwidth,keepaspectratio]{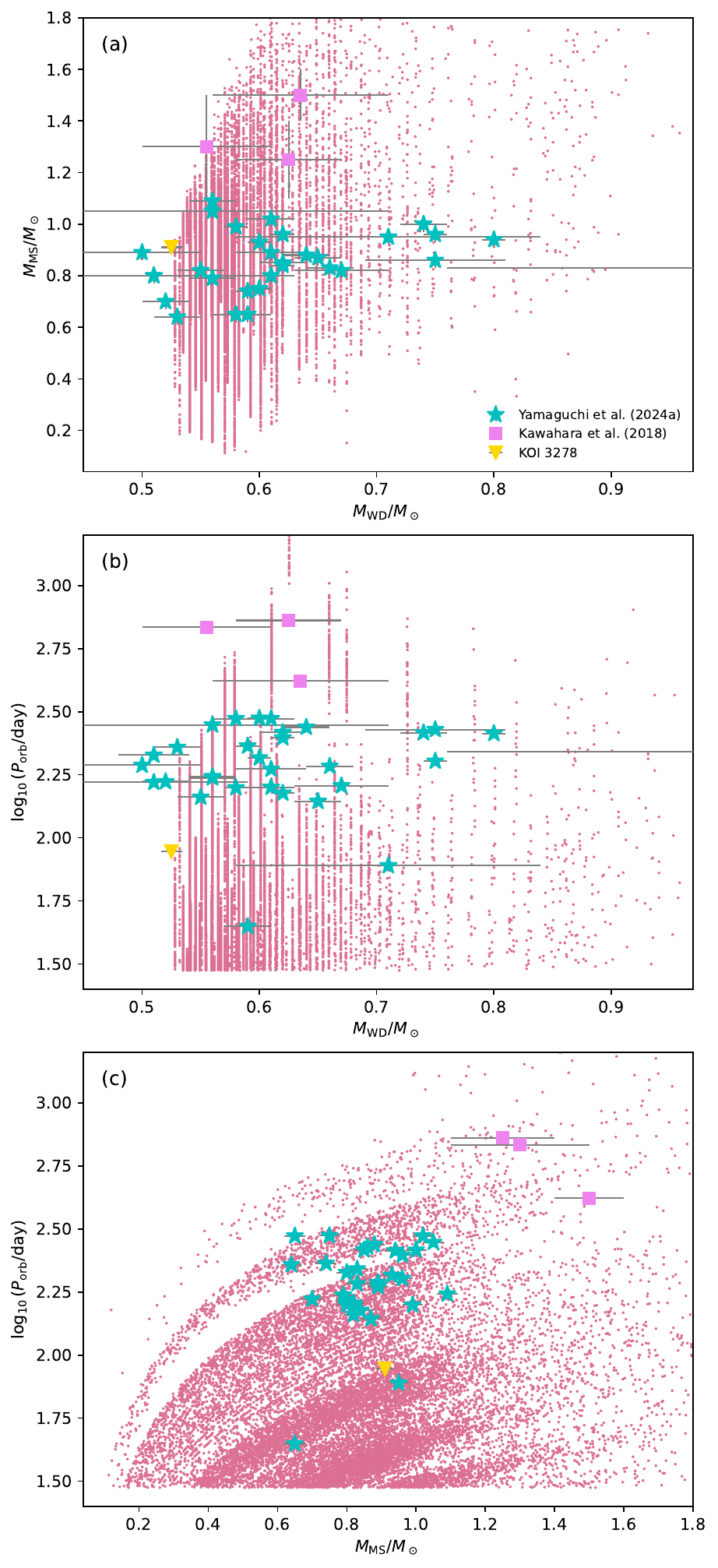}
\caption{Distributions of the long-period synthetic PCEBs (${P}_{\mathrm{orb}}>30$ days) obtained with ${\alpha}_{\mathrm{CE}}=1$ and ${\alpha}_{\mathrm{th}}=0$. Panels (a), (b), and (c) correspond to Figure~\ref{fig:population}(a), (c), and (e), respectively, with the same color coding and observational comparison.}
\label{fig:no_internal_energy}
\end{figure}

An important result is that long-period PCEBs with periods of several hundred days can form even without any contribution from internal energy ($({\alpha}_{\mathrm{CE}},{\alpha}_{\mathrm{th}})=(1,0)$, Figure~\ref{fig:no_internal_energy}). This is because highly evolved TP-AGB donors at thermal pulses can have envelopes weakly bound enough for orbital energy alone to eject them without substantial shrinkage. Thus, internal energy is not necessarily required to form wide PCEBs, provided that envelope ejection is sufficiently efficient.

Overall, our results show that long-period PCEBs are a general outcome of TP-AGB CEE across the explored CE parameter space, rather than being confined to a particular combination of ${\alpha}_{\mathrm{CE}}$ and ${\alpha}_{\mathrm{th}}$ (see Appendix~\ref{sec:appendix_ce} for the distributions of three additional parameter combinations). While these parameters strongly affect the relative abundance and maximum orbital period of the resulting systems, the underlying formation pathway remains viable even when the available envelope-ejection energy is considerably reduced.

\subsection{Initial--Final Mass Relation and Initial Parameter Space} \label{sec:ifmr}

\begin{figure}[ht!]
\centering
\includegraphics[width=\columnwidth, keepaspectratio]{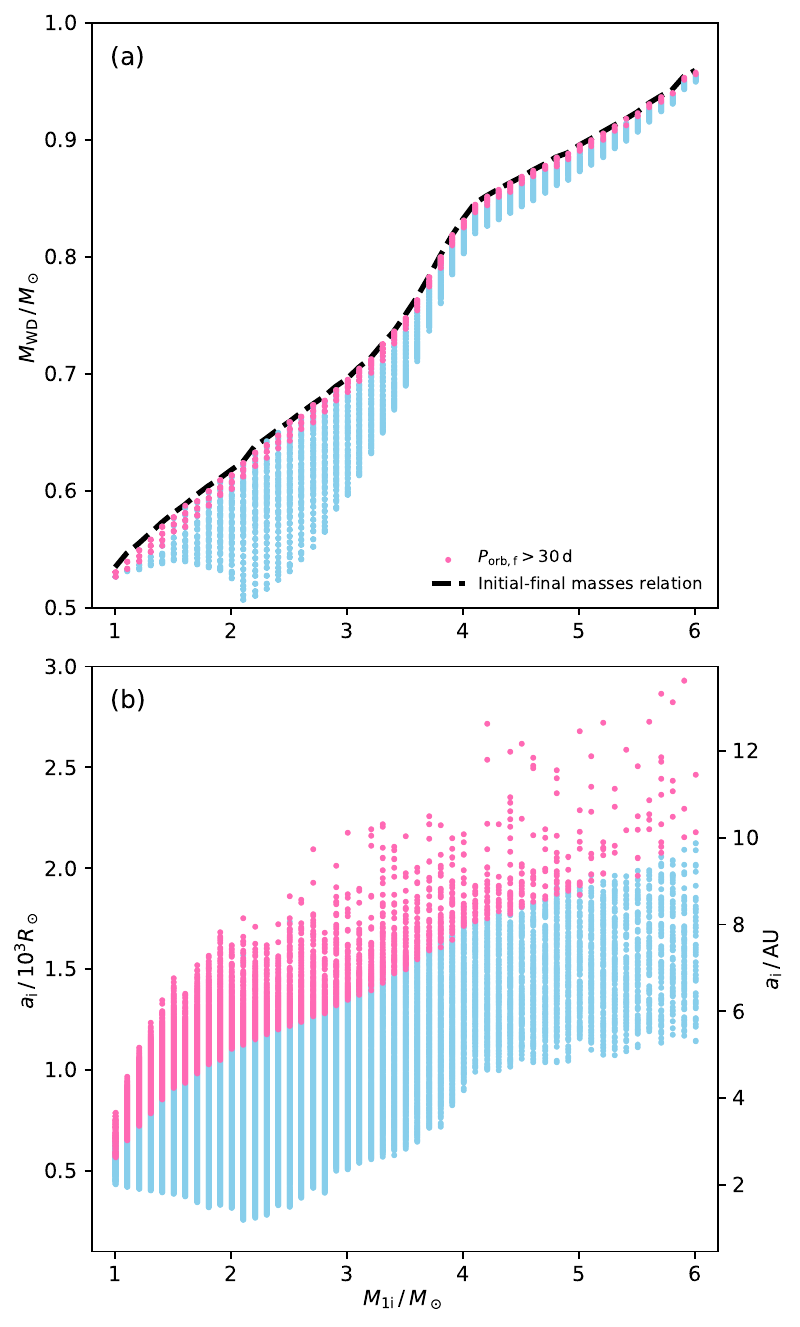}
\caption{(a) Initial primary mass versus final WD mass and (b) initial primary mass versus initial orbital separation for the synthetic PCEB population obtained with ${\alpha}_{\mathrm{CE}}=0.3$ and ${\alpha}_{\mathrm{th}}=0.5$. Pink points represent long-period systems (${P}_{\mathrm{orb}}>30$ days), while blue points indicate shorter-period systems (${P}_{\mathrm{orb}}\leq 30$ days). The black dashed line in panel (a) shows the single-star initial--final mass relation derived from the same MESA stellar evolution grid.}
\label{fig:ifmr}
\end{figure}

Figure~\ref{fig:ifmr}(a) compares the synthetic initial--final mass relation (IFMR) with the single-star IFMR from our MESA grid.  For a given initial primary mass, CEE can produce a range of WD masses because the core continues to grow during the TP-AGB phase, and envelope ejection terminates this growth at different evolutionary stages. The spread is particularly broad near $2\,M_\odot$, where TP-AGB core growth is most significant \citep{2014ApJ...782...17K}, and much narrower above approximately $4\,M_\odot$, where core growth is limited.

Interestingly, when we restrict the sample to long-period PCEBs, their WD masses closely follow the single-star IFMR across nearly the entire initial mass range. This behavior can be understood from the formation mechanism discussed in Section~\ref{sec:pulse_effects}. Long-period PCEBs frequently form when the donor undergoes CEE during the final few thermal pulses, after substantial wind mass loss has reduced the envelope binding energy. By this stage, the core has already accumulated most of its final mass, so subsequent envelope ejection produces a WD with a mass close to that expected from single-star evolution. In contrast, stable MT can occur over a broader range of evolutionary stages (specifically, the early AGB and early TP-AGB phases; \citealt{2026ApJ..1009...74Z}), and we therefore expect its initial--final masses to deviate from the single-star IFMR or have larger scatter. This difference may provide an observational diagnostic of formation channels, particularly in star clusters with independently constrained progenitor masses (Section~\ref{sec:discussion}).

Long-period PCEBs occupy a broad region of the initial primary mass--separation plane, extending to large separations (Figure~\ref{fig:ifmr}(b)). This is facilitated by pulse-driven expansion and tidal shrinkage, which promote RLOF in otherwise detached systems (Section~\ref{sec:pulse_effects}; \citealt{2016MNRAS.463.1040M}). Massive progenitors can produce long-period PCEBs over a wide separation range, but the widest generally require massive companions to initiate RLOF. Their short MS lifetimes strongly reduce the surviving population (Section~\ref{sec:population}), despite the broad accessible parameter space. Progenitors near $2\,M_\odot$ instead combine a broad formation region with long-lived low-mass companions and dominate the population, producing the WD-mass concentration near $0.6\,M_\odot$.

\section{Discussion} \label{sec:discussion}

In our models, CEE is predominantly triggered during thermal pulses, raising the question of whether the donor might contract before a CE interaction can fully develop. \citet{2022MNRAS.517.3181G,2024MNRAS.527.9145G} investigated this possibility using three-dimensional hydrodynamic simulations of a TP-AGB donor undergoing RLOF during the expansion phase of a thermal pulse. They found that the binary undergoes rapid orbital inspiral in less than about ten years, considerably shorter than the typical duration of thermal-pulse-driven envelope expansion, which can last several hundred years. This suggests that RLOF initiated during a thermal pulse can develop into CEE before the donor contracts again. However, whether similarly rapid inspiral occurs for highly evolved TP-AGB donors with very low envelope masses remains uncertain, particularly because these are the progenitors most relevant to the formation of long-period PCEBs in our models.

This naturally raises a more fundamental question: when a thermal pulse drives the donor into RLOF, what criteria govern the stability of MT? Unlike the relatively gradual expansion of RGB donors, thermal pulses can drive rapid envelope expansion over a much shorter timescale, while tidal interactions may simultaneously reduce the orbital separation (Section~\ref{sec:pulse_effects}; \citealt{2016MNRAS.463.1040M}). Even if an individual pulse does not initiate dynamically unstable MT, subsequent pulses with larger radius excursions may lead to stronger interactions \citep{2022MNRAS.517.3181G}. Moreover, \citet{2024PASP..136h4202Y} found that RLOF from TP-AGB donors can reach MT rates of approximately ${10}^{-2}M_\odot\,\mathrm{yr}^{-1}$, potentially allowing mass to escape through the outer Lagrangian points, thereby promoting unstable MT \citep{2020ApJS..249....9G}. These considerations suggest that thermal-pulse-driven RLOF may be particularly susceptible to instability, but its stability criteria have not yet been systematically established.

\begin{figure}[ht!]
\centering
\includegraphics[width=\columnwidth, keepaspectratio]{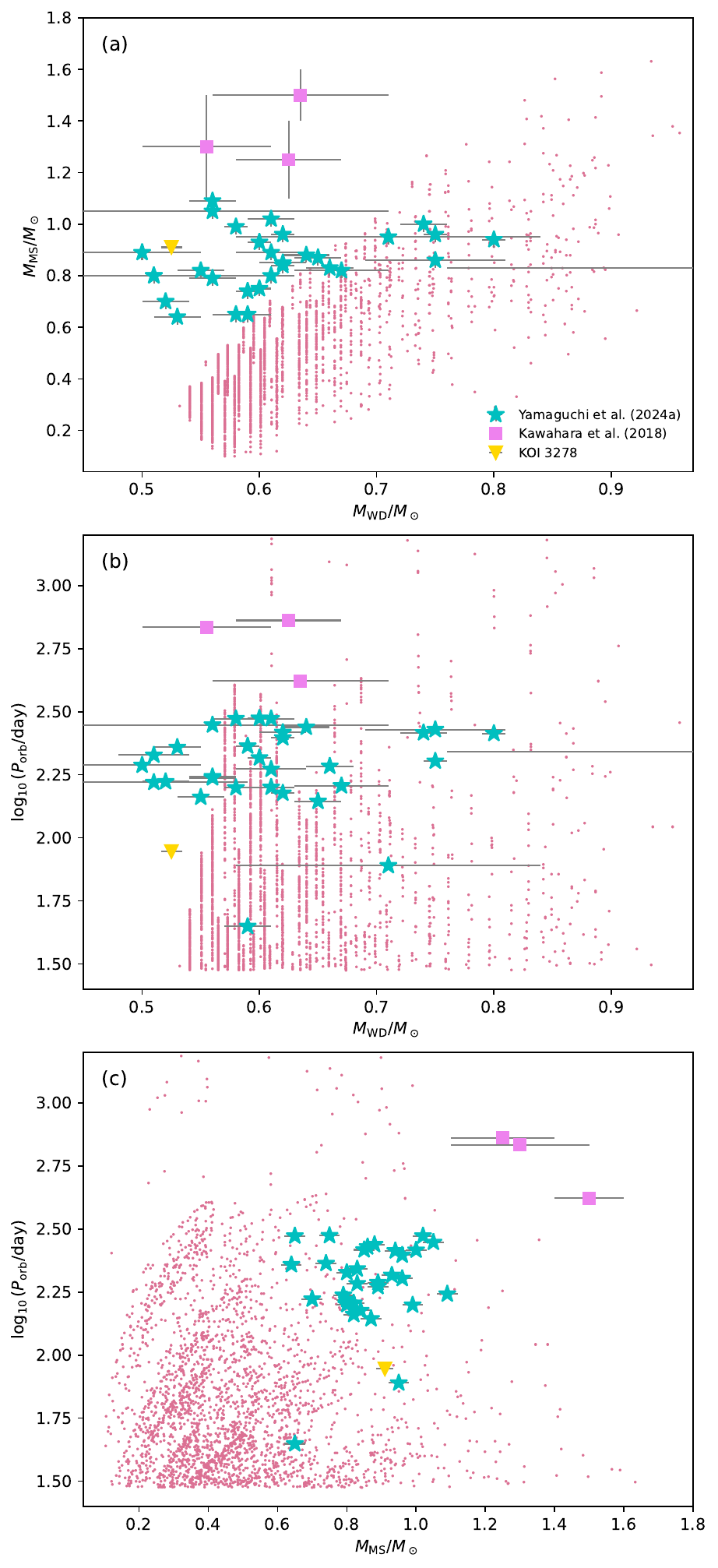}
\caption{Distributions of the long-period synthetic PCEBs (${P}_{\mathrm{orb}}>30$ days) obtained with ${\alpha}_{\mathrm{CE}}=0.3$, ${\alpha}_{\mathrm{th}}=0.5$, and an additional critical mass-ratio criterion of ${q}_{\mathrm{c}}=2$, such that CEE occurs only for ${M}_{\mathrm{d}}/{M}_{\mathrm{a}}>2$ at the onset of RLOF. The panel arrangement, color coding, and observational comparison are the same as in Figure~\ref{fig:no_internal_energy}.}
\label{fig:qcrit}
\end{figure}

In our BPS model, we therefore assume that RLOF initiated during the TP-AGB phase will directly lead to CEE. This assumption mainly considers the uncertainty in the stability criteria for thermal-pulse-driven MT and aims to explore a broader parameter space. However, to examine the dependence of our results, we also introduce a relatively strict critical mass-ratio criterion, ${q}_{\mathrm{c}}=2$, motivated by studies of MT stability \citep{2020ApJS..249....9G,2026ApJ..1007...15S}. Under this criterion, CEE is initiated only when the mass ratio at RLOF satisfies ${M}_{\mathrm{d}}/{M}_{\mathrm{a}}>2$. Figure~\ref{fig:qcrit} shows the resulting long-period PCEB population. Although this restriction substantially reduces the number of surviving systems, it does not prevent the formation of long-period PCEBs. Compared with the standard model in Figure~\ref{fig:population}, the remaining long-period systems predominantly contain low-mass MS companions. This is expected because late TP-AGB donors have already lost substantial envelope mass, making the condition ${M}_{\mathrm{d}}/{M}_{\mathrm{a}}>2$ increasingly difficult to satisfy unless the companion is relatively low in mass. Although their predicted abundance and companion-mass distribution change significantly, we expect that long-period systems with smaller companion masses can still form on a large scale via the CEE channel.

A further uncertainty concerns the physical evolution of TP-AGB stars near the end of their lives. As described in Section~\ref{sec:stellar_grid}, our MESA models employ numerical treatments to suppress the hydrogen recombination instability and Fe-peak instability, allowing the stars to evolve through the later TP-AGB phase. \citet{2024MNRAS.531.4033R} mentioned that the hydrogen recombination instability is likely a consequence of applying one-dimensional hydrostatic calculations to envelopes that are unstable to pulsations, whereas the Fe-peak instability may represent a genuine physical instability capable of driving substantial mass loss. In real TP-AGB stars, such instabilities may instead manifest as large-amplitude dynamical pulsations and rapid envelope ejection, which cannot be modeled by hydrostatic models. Indeed, \citet{2026A&A...707A.342C} performed one-dimensional hydrodynamic simulations of a TP-AGB star and found that decreasing envelope mass can lead to increasingly violent pulsations and substantial mass ejection near the end of the AGB phase. Related pulsation-driven mass loss has also been investigated in red supergiants \citep{2010ApJ...717L..62Y}. In this regime, the MT process may not be adequately described by classical stable MT or CEE formalism. Accurately understanding the associated mass and angular momentum evolution would therefore require time-dependent hydrodynamic modelling of the binary.

Throughout this work, we adopt ${\alpha}_{\mathrm{CE}}=0.3$ and ${\alpha}_{\mathrm{th}}=0.5$ as our standard model. The choice of ${\alpha}_{\mathrm{CE}}=0.3$ is motivated by numerous studies indicating that the majority of observed short-period PCEBs can be reproduced through relatively inefficient CEE \citep{2010A&A...520A..86Z,2013A&A...557A..87T,2014A&A...566A..86C,2017MNRAS.470.1442C,2022ApJ...933..137G,2024ApJ...961..202G,2025A&A...695A.161S,2026ApJ..1007...15S}. The appropriate efficiency for long-period systems, however, may be debatable. The contribution of internal energy to envelope ejection is also uncertain, and we adopt an intermediate value of ${\alpha}_{\mathrm{th}}=0.5$ for illustration. In particular, assuming that all internal energy can contribute to envelope ejection may be overly optimistic. Ionization/recombination energy constitutes an important component of the internal-energy reservoir in highly evolved TP-AGB envelopes \citep{2024PASP..136h4202Y}, but some of this energy may be released after substantial envelope expansion or transported away as radiation rather than converted into mechanical work \citep{2018MNRAS.478.1818G,2022MNRAS.516.4669L,2024MNRAS.528..234C}.

Another important result is that long-period PCEBs can still form without any internal-energy contribution when the CE ejection efficiency is sufficiently high (Section~\ref{sec:ce_parameters}). This is consistent with the broader findings of \citet{2024A&A...686A..61B} and \citet{2024PASP..136h4202Y}. However, \citet{2026ApJ..1010...73T} found that even when all thermal energy was included, long-period WD+MS binaries could not be explained without ionization/recombination energy, regardless of the adopted $\alpha_{\rm CE}$. We speculate that this difference may arise from the lack of self-consistency in traditional BPS models when resolving changes in stellar structure and binding energy across thermal pulses (see Section~\ref{sec:pulse_effects}). Whether differences in stellar structure, binding-energy prescriptions, or sample selection also contribute requires further investigation.

Star clusters offer a particularly effective test of the predicted agreement with the single-star IFMR (Section~\ref{sec:ifmr}). Cluster age, metallicity, and WD cooling age constrain the progenitor lifetime and initial mass, allowing comparison of the observed WD mass with the single-star IFMR. Open clusters are especially promising because their dynamics are generally less complex than those of globular clusters. \citet{2026arXiv260720611G} demonstrated this approach for a short-period PCEB in Alessi 12, whose massive WD is consistent with the single-star IFMR. This demonstrates that the approach is feasible, but testing our prediction calls for long-period WD+MS systems with progenitor masses near $2\,M_\odot$. Their substantial TP-AGB core growth produces the most pronounced distinction between short- and long-period PCEBs (Figure~\ref{fig:ifmr}(a)), making them particularly valuable for testing the predicted IFMR behavior and its use in distinguishing formation channels.

Our model also produces a substantial short-period PCEB population (Figure~\ref{fig:population}), suggesting the possibility of eventually modelling both short- and long-period systems within a unified framework, as explored by \citet{2026ApJ..1007...15S} and \citet{2026ApJ..1010...73T}. Detailed stellar structures could improve donor and binding-energy prescriptions across evolutionary phases, but comparison with the broader observed population requires extending beyond TP-AGB-initiated CEE to RGB and early AGB donors and initial masses above $6\,M_\odot$. Besides, nor do our results imply that all long-period WD+MS binaries originate from CEE. Stable MT from intermediate-mass donors can produce systems outside the classical WD mass--period relation \citep{2026ApJ..1009...74Z} and may explain some massive-companion systems reported by \citet{2018AJ....155..144K}. In our next study, we will incorporate stable MT to quantify the channels' relative contributions and formation parameter spaces, and predict a more complete long-period WD+MS population for systematic comparison with observations.

\section{Summary and Conclusion} \label{sec:summary}

We have developed a thermal-pulse-resolved BPS framework combining 51 detailed MESA tracks with prescriptions for detached wind mass loss and accretion, spin evolution, and tides. Applied to $5\times10^6$ binaries with initial primary masses of $1$--$6\,M_\odot$ at $Z=0.02$, it evaluates envelope binding energies from the stellar structure at RLOF and follows how individual pulses affect both interaction onset and CE outcome. Our main findings are:

\begin{enumerate}
\item Thermal pulses couple radius expansion, wind mass loss, and changes in moment of inertia to the wind- and tide-driven orbital evolution. Near Roche-lobe filling, pulse-driven expansion and tidal shrinkage can jointly promote RLOF. The weakly bound envelopes of highly evolved TP-AGB donors then allow CEE to leave relatively wide post-CE orbits.

\item The population spans a wide range of orbital periods, including hundreds of days. In the standard model ($\alpha_{\mathrm{CE}}=0.3$, $\alpha_{\mathrm{th}}=0.5$), approximately 11.5\% of TP-AGB-initiated PCEBs have $P_{\mathrm{orb}}>30$ days. Long-period systems form for all 16 CE parameter combinations, with fractions of 0.77\%--42.89\%. At $\alpha_{\rm CE}=1$, periods of several hundred days are possible without internal energy, which is therefore not a necessary condition for long-period PCEBs in this framework.

\item WDs near $0.6\,M_\odot$ dominate the long-period population, with few massive WDs. More massive progenitors generally have more strongly bound envelopes, requiring more orbital energy to retain wide post-CE orbits, while the more massive companions supplying this energy have shorter MS lifetimes. Long-period PCEBs preferentially emerge from the late TP-AGB, after most core growth, and thus closely follow the single-star IFMR. This provides a testable prediction for open-cluster WD+MS binaries with independently constrained progenitor masses.

\end{enumerate}

Resolving thermal pulses and their structural changes provides a physically motivated route to long-period PCEB formation through CEE. This channel's contribution to the Gaia population remains linked to TP-AGB MT stability and envelope-ejection physics. Extending the framework to stable MT and other giant-branch phases will enable a more complete comparison of formation channels.

\begin{acknowledgments}

This project is supported by the National Natural Science Foundation of China (NSFC Nos. 12525304, 12288102, 12125303, W2632018, 12473033, 12333008 and 12422305), the National Key R\&D Program of China (No. 2021YFA1600403), the Strategic Priority Research Program of the Chinese Academy of Sciences (grant No. XDB1160201), the Yunnan Revitalization Talent Support Program—Science \& Technology Champion Project (No. 202305AB350003), the Yunnan Fundamental Research Projects (No. 202401BC070007, 202601CJ070008), the Young Talent Project of Yunnan Revitalization Talent Support Program, the International Centre of Supernovae, Yunnan Key Laboratory (No. 202302AN360001).

\end{acknowledgments}

\appendix
\setcounter{figure}{0}
\renewcommand{\thefigure}{\Alph{section}\arabic{figure}}       % 显示：A1、A2、A3
\renewcommand{\theHfigure}{\Alph{section}\arabic{figure}}

\section{PCEB Distributions for Additional CE Parameter Combinations}
\label{sec:appendix_ce}

\begin{figure*}[ht!]
\centering
\includegraphics[width=\textwidth]{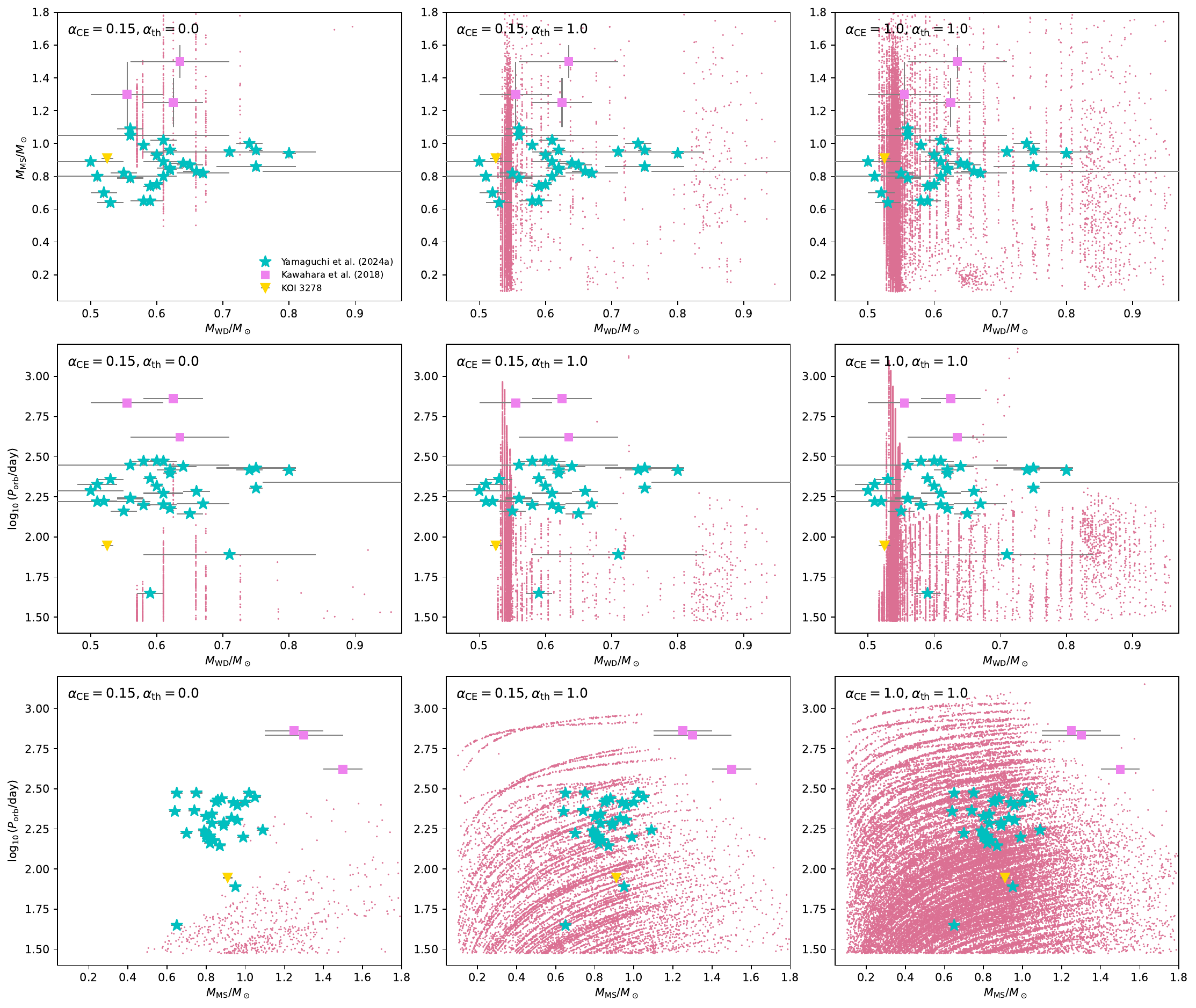}
\caption{Distributions of the long-period synthetic PCEBs (${P}_{\mathrm{orb}}>30$ days) for $(\alpha_{\rm CE},\alpha_{\rm th})=(0.15,0)$, $(0.15,1)$, and $(1,1)$, from left to right. The top, middle, and bottom rows show companion mass versus WD mass, orbital period versus WD mass, and orbital period versus companion mass, respectively. The color coding and observational comparison are the same as in the long-period panels of Figure~\ref{fig:no_internal_energy}.}
\label{fig:additional_ce}
\end{figure*}

Figure~\ref{fig:additional_ce} presents the long-period synthetic PCEBs for three additional extreme combinations of CE parameters sampled in Figure~\ref{fig:ce_parameters}: $(\alpha_{\rm CE},\alpha_{\rm th})=(0.15,0)$, $(0.15,1)$, and $(1,1)$. These complement the standard model and the $(\alpha_{\rm CE},\alpha_{\rm th})=(1,0)$ model presented in the main text.

\bibliography{sample702}

@ARTICLE{2013ARA&A..51..269D,
       author = {{Duch{\^e}ne}, Gaspard and {Kraus}, Adam},
        title = "{Stellar Multiplicity}",
      journal = {\araa},
         year = 2013,
        month = aug,
       volume = {51},
       number = {1},
        pages = {269-310},
          doi = {10.1146/annurev-astro-081710-102602},
archivePrefix = {arXiv},
       eprint = {1303.3028},
 primaryClass = {astro-ph.SR},
       adsurl = {https://ui.adsabs.harvard.edu/abs/2013ARA&A..51..269D}
}

@ARTICLE{2017ApJS..230...15M,
       author = {{Moe}, Maxwell and {Di Stefano}, Rosanne},
        title = "{Mind Your Ps and Qs: The Interrelation between Period (P) and Mass-ratio (Q) Distributions of Binary Stars}",
      journal = {\apjs},
         year = 2017,
        month = jun,
       volume = {230},
       number = {2},
          eid = {15},
        pages = {15},
          doi = {10.3847/1538-4365/aa6fb6},
archivePrefix = {arXiv},
       eprint = {1606.05347},
 primaryClass = {astro-ph.SR},
       adsurl = {https://ui.adsabs.harvard.edu/abs/2017ApJS..230...15M}
}

@ARTICLE{2024PrPNP.13404083C,
       author = {{Chen}, Xuefei and {Liu}, Zhengwei and {Han}, Zhanwen},
        title = "{Binary stars in the new millennium}",
      journal = {Progress in Particle and Nuclear Physics},
         year = 2024,
        month = jan,
       volume = {134},
          eid = {104083},
        pages = {104083},
          doi = {10.1016/j.ppnp.2023.104083},
archivePrefix = {arXiv},
       eprint = {2311.11454},
 primaryClass = {astro-ph.SR},
       adsurl = {https://ui.adsabs.harvard.edu/abs/2024PrPNP.13404083C}
}

@ARTICLE{1984ApJ...277..355W,
       author = {{Webbink}, R.~F.},
        title = "{Double white dwarfs as progenitors of R Coronae Borealis stars and type I supernovae.}",
      journal = {\apj},
         year = 1984,
        month = feb,
       volume = {277},
        pages = {355-360},
          doi = {10.1086/161701},
       adsurl = {https://ui.adsabs.harvard.edu/abs/1984ApJ...277..355W}
}

@ARTICLE{2004MNRAS.350.1301H,
       author = {{Han}, Z. and {Podsiadlowski}, Ph.},
        title = "{The single-degenerate channel for the progenitors of Type Ia supernovae}",
      journal = {\mnras},
         year = 2004,
        month = jun,
       volume = {350},
       number = {4},
        pages = {1301-1309},
          doi = {10.1111/j.1365-2966.2004.07713.x},
archivePrefix = {arXiv},
       eprint = {astro-ph/0309618},
 primaryClass = {astro-ph},
       adsurl = {https://ui.adsabs.harvard.edu/abs/2004MNRAS.350.1301H}
}

@ARTICLE{1987ApJ...314..653W,
       author = {{Webbink}, Ronald F. and {Livio}, Mario and {Truran}, James W. and {Orio}, Marina},
        title = "{The Nature of the Recurrent Novae}",
      journal = {\apj},
         year = 1987,
        month = mar,
       volume = {314},
        pages = {653},
          doi = {10.1086/165095},
       adsurl = {https://ui.adsabs.harvard.edu/abs/1987ApJ...314..653W}
}

@ARTICLE{2003A&A...406..305S,
       author = {{Schreiber}, M.~R. and {G{\"a}nsicke}, B.~T.},
        title = "{The age, life expectancy, and space density   of Post Common Envelope Binaries}",
      journal = {\aap},
         year = 2003,
        month = jul,
       volume = {406},
        pages = {305-321},
          doi = {10.1051/0004-6361:20030801},
archivePrefix = {arXiv},
       eprint = {astro-ph/0305531},
 primaryClass = {astro-ph},
       adsurl = {https://ui.adsabs.harvard.edu/abs/2003A&A...406..305S}
}

@ARTICLE{1995ApJ...447..656Y,
       author = {{Yungelson}, L. and {Livio}, M. and {Tutukov}, A. and {Kenyon}, S.~J.},
        title = "{A Model for the Galactic Population of Symbiotic Stars with White Dwarf Accretors}",
      journal = {\apj},
         year = 1995,
        month = jul,
       volume = {447},
        pages = {656},
          doi = {10.1086/175908},
       adsurl = {https://ui.adsabs.harvard.edu/abs/1995ApJ...447..656Y}
}

@ARTICLE{2024A&A...686A.226B,
       author = {{Belloni}, Diogo and {Miko{\l}ajewska}, Joanna and {Schreiber}, Matthias R.},
        title = "{The formation of the magnetic symbiotic star FN Sgr}",
      journal = {\aap},
         year = 2024,
        month = jun,
       volume = {686},
          eid = {A226},
        pages = {A226},
          doi = {10.1051/0004-6361/202449602},
archivePrefix = {arXiv},
       eprint = {2402.08647},
 primaryClass = {astro-ph.SR},
       adsurl = {https://ui.adsabs.harvard.edu/abs/2024A&A...686A.226B}
}

@ARTICLE{2003MNRAS.340.1214P,
       author = {{Podsiadlowski}, Ph. and {Han}, Z. and {Rappaport}, S.},
        title = "{Cataclysmic variables with evolved secondaries and the progenitors of AM CVn stars}",
      journal = {\mnras},
         year = 2003,
        month = apr,
       volume = {340},
       number = {4},
        pages = {1214-1228},
          doi = {10.1046/j.1365-8711.2003.06380.x},
       adsurl = {https://ui.adsabs.harvard.edu/abs/2003MNRAS.340.1214P}
}

@ARTICLE{2001A&A...375..890N,
       author = {{Nelemans}, G. and {Yungelson}, L.~R. and {Portegies Zwart}, S.~F.},
        title = "{The gravitational wave signal from the Galactic disk population of binaries containing two compact objects}",
      journal = {\aap},
         year = 2001,
        month = sep,
       volume = {375},
        pages = {890-898},
          doi = {10.1051/0004-6361:20010683},
archivePrefix = {arXiv},
       eprint = {astro-ph/0105221},
 primaryClass = {astro-ph},
       adsurl = {https://ui.adsabs.harvard.edu/abs/2001A&A...375..890N}
}

@ARTICLE{2004A&A...419.1057W,
       author = {{Willems}, B. and {Kolb}, U.},
        title = "{Detached white dwarf main-sequence star binaries}",
      journal = {\aap},
         year = 2004,
        month = jun,
       volume = {419},
        pages = {1057-1076},
          doi = {10.1051/0004-6361:20040085},
archivePrefix = {arXiv},
       eprint = {astro-ph/0403090},
 primaryClass = {astro-ph},
       adsurl = {https://ui.adsabs.harvard.edu/abs/2004A&A...419.1057W}
}

@INPROCEEDINGS{1976IAUS...73...75P,
       author = {{Paczynski}, B.},
        title = "{Common Envelope Binaries}",
    booktitle = {Structure and Evolution of Close Binary Systems},
         year = 1976,
       editor = {{Eggleton}, Peter and {Mitton}, Simon and {Whelan}, John},
       series = {IAU Symposium},
       volume = {73},
        month = jan,
        pages = {75},
       adsurl = {https://ui.adsabs.harvard.edu/abs/1976IAUS...73...75P}
}

@PHDTHESIS{1975PhDT.......165W,
       author = {{Webbink}, R.~F.},
        title = "{The evolution of low-mass close binary systems}",
       school = {University of Cambridge, UK},
         year = 1975,
        month = jul,
       adsurl = {https://ui.adsabs.harvard.edu/abs/1975PhDT.......165W}
}

@ARTICLE{2014A&A...566A..86C,
       author = {{Camacho}, Judit and {Torres}, Santiago and {Garc{\'\i}a-Berro}, Enrique and {Zorotovic}, M{\'o}nica and {Schreiber}, Matthias R. and {Rebassa-Mansergas}, Alberto and {Nebot G{\'o}mez-Mor{\'a}n}, Ada and {G{\"a}nsicke}, Boris T.},
        title = "{Monte Carlo simulations of post-common-envelope white dwarf + main sequence binaries: comparison with the SDSS DR7 observed sample}",
      journal = {\aap},
         year = 2014,
        month = jun,
       volume = {566},
          eid = {A86},
        pages = {A86},
          doi = {10.1051/0004-6361/201323052},
archivePrefix = {arXiv},
       eprint = {1404.5464},
 primaryClass = {astro-ph.GA},
       adsurl = {https://ui.adsabs.harvard.edu/abs/2014A&A...566A..86C}
}

@ARTICLE{2026ApJ..1007...15S,
       author = {{Shi}, Yanxu and {Ge}, Hongwei and {Li}, Zhenwei and {Belloni}, Diogo and {Zheng}, Rizhong and {Jiang}, Dengkai and {Chen}, Hailiang and {Santos-Garcia}, A. and {Torres}, S. and {Rebassa-Mansergas}, A. and {Chen}, Xuefei and {Han}, Zhanwen},
        title = "{Influence of Mass-transfer Stability on the Formation of Post-common-envelope Binaries}",
      journal = {\apj},
         year = 2026,
        month = aug,
       volume = {1007},
       number = {1},
          eid = {15},
        pages = {15},
          doi = {10.3847/1538-4357/ae876a},
archivePrefix = {arXiv},
       eprint = {2606.22927},
 primaryClass = {astro-ph.SR},
       adsurl = {https://ui.adsabs.harvard.edu/abs/2026ApJ..1007...15S}
}

@ARTICLE{1995MNRAS.273..731R,
       author = {{Rappaport}, S. and {Podsiadlowski}, Ph. and {Joss}, P.~C. and {Di Stefano}, R. and {Han}, Z.},
        title = "{The relation between white dwarf mass and orbital period in wide binary radio pulsars}",
      journal = {\mnras},
         year = 1995,
        month = apr,
       volume = {273},
       number = {3},
        pages = {731-741},
          doi = {10.1093/mnras/273.3.731},
       adsurl = {https://ui.adsabs.harvard.edu/abs/1995MNRAS.273..731R}
}

@ARTICLE{1999A&A...350..928T,
       author = {{Tauris}, Thomas M. and {Savonije}, Gerrit J.},
        title = "{Formation of millisecond pulsars. I. Evolution of low-mass X-ray binaries with P\_orb> 2 days}",
      journal = {\aap},
         year = 1999,
        month = oct,
       volume = {350},
        pages = {928-944},
          doi = {10.48550/arXiv.astro-ph/9909147},
archivePrefix = {arXiv},
       eprint = {astro-ph/9909147},
 primaryClass = {astro-ph},
       adsurl = {https://ui.adsabs.harvard.edu/abs/1999A&A...350..928T}
}

@ARTICLE{2014MNRAS.437.2217S,
       author = {{Smedley}, Sarah L. and {Tout}, Christopher A. and {Ferrario}, Lilia and {Wickramasinghe}, Dayal T.},
        title = "{The nature of millisecond pulsars with helium white dwarf companions}",
      journal = {\mnras},
         year = 2014,
        month = jan,
       volume = {437},
       number = {3},
        pages = {2217-2229},
          doi = {10.1093/mnras/stt2030},
archivePrefix = {arXiv},
       eprint = {1310.8356},
 primaryClass = {astro-ph.SR},
       adsurl = {https://ui.adsabs.harvard.edu/abs/2014MNRAS.437.2217S}
}

@ARTICLE{2026ApJ...996...74M,
       author = {{Mou}, Jian and {Chen}, Hai-Liang and {Jiang}, Dengkai and {Ge}, Hongwei and {Zhang}, Lifu and {Zheng}, Rizhong and {Chen}, Xuefei and {Han}, Zhanwen},
        title = "{The Relation between Helium White Dwarf Mass and Orbital Period under Two Types of Opacity}",
      journal = {\apj},
         year = 2026,
        month = jan,
       volume = {996},
       number = {1},
          eid = {74},
        pages = {74},
          doi = {10.3847/1538-4357/ae23d1},
archivePrefix = {arXiv},
       eprint = {2511.20147},
 primaryClass = {astro-ph.SR},
       adsurl = {https://ui.adsabs.harvard.edu/abs/2026ApJ...996...74M}
}

@ARTICLE{2026ApJ..1009...74Z,
       author = {{Zheng}, Rizhong and {Ge}, Hongwei and {Tout}, Christopher A. and {Chen}, Hailiang and {Li}, Zhenwei and {Jiang}, Dengkai and {Li}, Chengyuan and {Tian}, Zhijia and {Ma}, Bo and {Zhang}, Lifu and {Mou}, Jian and {Chen}, Xuefei and {Han}, Zhanwen},
        title = "{Mass-Orbital Period Distribution of Massive White Dwarfs Formed through Stable Mass Transfer}",
      journal = {\apj},
         year = 2026,
        month = sep,
       volume = {1009},
       number = {1},
          eid = {74},
        pages = {74},
          doi = {10.3847/1538-4357/ae9d74},
archivePrefix = {arXiv},
       eprint = {2606.06141},
 primaryClass = {astro-ph.SR},
       adsurl = {https://ui.adsabs.harvard.edu/abs/2026ApJ..1009...74Z}
}

@ARTICLE{2019MNRAS.484.5362A,
       author = {{Ashley}, R.~P. and {Farihi}, J. and {Marsh}, T.~R. and {Wilson}, D.~J. and {G{\"a}nsicke}, B.~T.},
        title = "{Evidence for bimodal orbital separations of white dwarf-red dwarf binary stars}",
      journal = {\mnras},
         year = 2019,
        month = apr,
       volume = {484},
       number = {4},
        pages = {5362-5376},
          doi = {10.1093/mnras/stz298},
archivePrefix = {arXiv},
       eprint = {1901.09139},
 primaryClass = {astro-ph.SR},
       adsurl = {https://ui.adsabs.harvard.edu/abs/2019MNRAS.484.5362A}
}

@ARTICLE{2022MNRAS.512.2625L,
       author = {{Lagos}, F. and {Schreiber}, M.~R. and {Parsons}, S.~G. and {Toloza}, O. and {G{\"a}nsicke}, B.~T. and {Hernandez}, M.~S. and {Schmidtobreick}, L. and {Belloni}, D.},
        title = "{The white dwarf binary pathways survey - VII. Evidence for a bi-modal distribution of post-mass transfer systems?}",
      journal = {\mnras},
         year = 2022,
        month = may,
       volume = {512},
       number = {2},
        pages = {2625-2635},
          doi = {10.1093/mnras/stac673},
archivePrefix = {arXiv},
       eprint = {2203.08846},
 primaryClass = {astro-ph.SR},
       adsurl = {https://ui.adsabs.harvard.edu/abs/2022MNRAS.512.2625L}
}

@ARTICLE{2024MNRAS.529.3729S,
       author = {{Shahaf}, S. and {Hallakoun}, N. and {Mazeh}, T. and {Ben-Ami}, S. and {Rekhi}, P. and {El-Badry}, K. and {Toonen}, S.},
        title = "{Triage of the Gaia DR3 astrometric orbits. II. A census of white dwarfs}",
      journal = {\mnras},
         year = 2024,
        month = apr,
       volume = {529},
       number = {4},
        pages = {3729-3743},
          doi = {10.1093/mnras/stae773},
archivePrefix = {arXiv},
       eprint = {2309.15143},
 primaryClass = {astro-ph.SR},
       adsurl = {https://ui.adsabs.harvard.edu/abs/2024MNRAS.529.3729S}
}

@ARTICLE{2024MNRAS.52711719Y,
       author = {{Yamaguchi}, Natsuko and {El-Badry}, Kareem and {Fuller}, Jim and {Latham}, David W. and {Cargile}, Phillip A. and {Mazeh}, Tsevi and {Shahaf}, Sahar and {Bieryla}, Allyson and {Buchhave}, Lars A. and {Hobson}, Melissa},
        title = "{Wide post-common envelope binaries containing ultramassive white dwarfs: evidence for efficient envelope ejection in massive asymptotic giant branch stars}",
      journal = {\mnras},
         year = 2024,
        month = feb,
       volume = {527},
       number = {4},
        pages = {11719-11739},
          doi = {10.1093/mnras/stad4005},
archivePrefix = {arXiv},
       eprint = {2309.15905},
 primaryClass = {astro-ph.SR},
       adsurl = {https://ui.adsabs.harvard.edu/abs/2024MNRAS.52711719Y}
}

@ARTICLE{2024PASP..136h4202Y,
       author = {{Yamaguchi}, Natsuko and {El-Badry}, Kareem and {Rees}, Natalie R. and {Shahaf}, Sahar and {Mazeh}, Tsevi and {Andrae}, Ren{\'e}},
        title = "{Wide Post-common Envelope Binaries from Gaia: Orbit Validation and Formation Models}",
      journal = {\pasp},
         year = 2024,
        month = aug,
       volume = {136},
       number = {8},
          eid = {084202},
        pages = {084202},
          doi = {10.1088/1538-3873/ad6809},
archivePrefix = {arXiv},
       eprint = {2405.06020},
 primaryClass = {astro-ph.SR},
       adsurl = {https://ui.adsabs.harvard.edu/abs/2024PASP..136h4202Y}
}

@ARTICLE{2025A&A...704A...6R,
       author = {{Rubio}, A.~C. and {Breivik}, K. and {Badenes}, C. and {El-Badry}, K. and {Anguiano}, B. and {Linck}, E. and {Majewski}, S. and {Stassun}, K.},
        title = "{Calibration of binary population synthesis models using white dwarf binaries from APOGEE, GALEX, and Gaia}",
      journal = {\aap},
         year = 2025,
        month = nov,
       volume = {704},
          eid = {A6},
        pages = {A6},
          doi = {10.1051/0004-6361/202555600},
archivePrefix = {arXiv},
       eprint = {2505.15893},
 primaryClass = {astro-ph.SR},
       adsurl = {https://ui.adsabs.harvard.edu/abs/2025A&A...704A...6R}
}

@ARTICLE{2026ApJS..286...13V,
       author = {{van Son}, Lieke A.~C. and {Yamaguchi}, Natsuko and {Nagarajan}, Pranav and {Shenar}, Tomer and {Sen}, Koushik and {Laroche}, Alexander and {Leiner}, Emily M. and {Sana}, Hugues and {Pols}, Onno R.},
        title = "{Ongoing and Post-mass-transfer Binaries: A Living Catalog and Unified Review of Binary Mass-transfer Products}",
      journal = {\apjs},
         year = 2026,
        month = sep,
       volume = {286},
       number = {1},
          eid = {13},
        pages = {13},
          doi = {10.3847/1538-4365/ae8d04},
archivePrefix = {arXiv},
       eprint = {2605.31290},
 primaryClass = {astro-ph.SR},
       adsurl = {https://ui.adsabs.harvard.edu/abs/2026ApJS..286...13V}
}

@ARTICLE{2022MNRAS.517.3181G,
       author = {{Gonz{\'a}lez-Bol{\'\i}var}, Miguel and {De Marco}, Orsola and {Lau}, Mike Y.~M. and {Hirai}, Ryosuke and {Price}, Daniel J.},
        title = "{Common envelope binary interaction simulations between a thermally pulsating AGB star and a low mass companion}",
      journal = {\mnras},
         year = 2022,
        month = dec,
       volume = {517},
       number = {3},
        pages = {3181-3199},
          doi = {10.1093/mnras/stac2301},
archivePrefix = {arXiv},
       eprint = {2205.09749},
 primaryClass = {astro-ph.SR},
       adsurl = {https://ui.adsabs.harvard.edu/abs/2022MNRAS.517.3181G}
}

@ARTICLE{2024MNRAS.527.9145G,
       author = {{Gonz{\'a}lez-Bol{\'\i}var}, Miguel and {De Marco}, Orsola and {Berm{\'u}dez-Bustamante}, Luis C. and {Siess}, Lionel and {Price}, Daniel J.},
        title = "{Dust formation in common envelope binary interaction - I: 3D simulations using the Bowen approximation}",
      journal = {\mnras},
         year = 2024,
        month = jan,
       volume = {527},
       number = {3},
        pages = {9145-9158},
          doi = {10.1093/mnras/stad3748},
archivePrefix = {arXiv},
       eprint = {2306.16609},
 primaryClass = {astro-ph.SR},
       adsurl = {https://ui.adsabs.harvard.edu/abs/2024MNRAS.527.9145G}
}

@ARTICLE{2024A&A...686A..61B,
       author = {{Belloni}, Diogo and {Zorotovic}, Monica and {Schreiber}, Matthias R. and {Parsons}, Steven G. and {Moe}, Maxwell and {Garbutt}, James A.},
        title = "{Formation of long-period post-common envelope binaries. I. No extra energy is needed to explain oxygen-neon white dwarfs paired with AFGK-type main-sequence stars}",
      journal = {\aap},
         year = 2024,
        month = jun,
       volume = {686},
          eid = {A61},
        pages = {A61},
          doi = {10.1051/0004-6361/202449235},
archivePrefix = {arXiv},
       eprint = {2401.07692},
 primaryClass = {astro-ph.SR},
       adsurl = {https://ui.adsabs.harvard.edu/abs/2024A&A...686A..61B}
}

@ARTICLE{2024A&A...687A..12B,
       author = {{Belloni}, Diogo and {Schreiber}, Matthias R. and {Zorotovic}, Monica},
        title = "{Formation of long-period post-common-envelope binaries. II. Explaining the self-lensing binary KOI 3278}",
      journal = {\aap},
         year = 2024,
        month = jul,
       volume = {687},
          eid = {A12},
        pages = {A12},
          doi = {10.1051/0004-6361/202449320},
archivePrefix = {arXiv},
       eprint = {2401.17510},
 primaryClass = {astro-ph.SR},
       adsurl = {https://ui.adsabs.harvard.edu/abs/2024A&A...687A..12B}
}

@ARTICLE{2000MNRAS.315..543H,
       author = {{Hurley}, Jarrod R. and {Pols}, Onno R. and {Tout}, Christopher A.},
        title = "{Comprehensive analytic formulae for stellar evolution as a function of mass and metallicity}",
      journal = {\mnras},
         year = 2000,
        month = jul,
       volume = {315},
       number = {3},
        pages = {543-569},
          doi = {10.1046/j.1365-8711.2000.03426.x},
archivePrefix = {arXiv},
       eprint = {astro-ph/0001295},
 primaryClass = {astro-ph},
       adsurl = {https://ui.adsabs.harvard.edu/abs/2000MNRAS.315..543H}
}

@ARTICLE{2002MNRAS.329..897H,
       author = {{Hurley}, Jarrod R. and {Tout}, Christopher A. and {Pols}, Onno R.},
        title = "{Evolution of binary stars and the effect of tides on binary populations}",
      journal = {\mnras},
         year = 2002,
        month = feb,
       volume = {329},
       number = {4},
        pages = {897-928},
          doi = {10.1046/j.1365-8711.2002.05038.x},
archivePrefix = {arXiv},
       eprint = {astro-ph/0201220},
 primaryClass = {astro-ph},
       adsurl = {https://ui.adsabs.harvard.edu/abs/2002MNRAS.329..897H}
}

@ARTICLE{2020ApJ...898...71B,
       author = {{Breivik}, Katelyn and {Coughlin}, Scott and {Zevin}, Michael and {Rodriguez}, Carl L. and {Kremer}, Kyle and {Ye}, Claire S. and {Andrews}, Jeff J. and {Kurkowski}, Michael and {Digman}, Matthew C. and {Larson}, Shane L. and {Rasio}, Frederic A.},
        title = "{COSMIC Variance in Binary Population Synthesis}",
      journal = {\apj},
         year = 2020,
        month = jul,
       volume = {898},
       number = {1},
          eid = {71},
        pages = {71},
          doi = {10.3847/1538-4357/ab9d85},
archivePrefix = {arXiv},
       eprint = {1911.00903},
 primaryClass = {astro-ph.HE},
       adsurl = {https://ui.adsabs.harvard.edu/abs/2020ApJ...898...71B}
}

@ARTICLE{2022ApJS..258...34R,
       author = {{Riley}, Jeff and {Agrawal}, Poojan and {Barrett}, Jim W. and {Boyett}, Kristan N.~K. and {Broekgaarden}, Floor S. and {Chattopadhyay}, Debatri and {Gaebel}, Sebastian M. and {Gittins}, Fabian and {Hirai}, Ryosuke and {Howitt}, George and {Justham}, Stephen and {Khandelwal}, Lokesh and {Kummer}, Floris and {Lau}, Mike Y.~M. and {Mandel}, Ilya and {de Mink}, Selma E. and {Neijssel}, Coenraad and {Riley}, Tim and {van Son}, Lieke and {Stevenson}, Simon and {Vigna-G{\'o}mez}, Alejandro and {Vinciguerra}, Serena and {Wagg}, Tom and {Willcox}, Reinhold and {Team Compas}},
        title = "{Rapid Stellar and Binary Population Synthesis with COMPAS}",
      journal = {\apjs},
         year = 2022,
        month = feb,
       volume = {258},
       number = {2},
          eid = {34},
        pages = {34},
          doi = {10.3847/1538-4365/ac416c},
archivePrefix = {arXiv},
       eprint = {2109.10352},
 primaryClass = {astro-ph.IM},
       adsurl = {https://ui.adsabs.harvard.edu/abs/2022ApJS..258...34R}
}

@ARTICLE{2004MNRAS.350..407I,
       author = {{Izzard}, Robert G. and {Tout}, Christopher A. and {Karakas}, Amanda I. and {Pols}, Onno R.},
        title = "{A new synthetic model for asymptotic giant branch stars}",
      journal = {\mnras},
         year = 2004,
        month = may,
       volume = {350},
       number = {2},
        pages = {407-426},
          doi = {10.1111/j.1365-2966.2004.07446.x},
archivePrefix = {arXiv},
       eprint = {astro-ph/0402403},
 primaryClass = {astro-ph},
       adsurl = {https://ui.adsabs.harvard.edu/abs/2004MNRAS.350..407I}
}

@ARTICLE{2011ApJS..192....3P,
       author = {{Paxton}, Bill and {Bildsten}, Lars and {Dotter}, Aaron and {Herwig}, Falk and {Lesaffre}, Pierre and {Timmes}, Frank},
        title = "{Modules for Experiments in Stellar Astrophysics (MESA)}",
      journal = {\apjs},
         year = 2011,
        month = jan,
       volume = {192},
       number = {1},
          eid = {3},
        pages = {3},
          doi = {10.1088/0067-0049/192/1/3},
archivePrefix = {arXiv},
       eprint = {1009.1622},
 primaryClass = {astro-ph.SR},
       adsurl = {https://ui.adsabs.harvard.edu/abs/2011ApJS..192....3P}
}

@ARTICLE{2013ApJS..208....4P,
       author = {{Paxton}, Bill and {Cantiello}, Matteo and {Arras}, Phil and {Bildsten}, Lars and {Brown}, Edward F. and {Dotter}, Aaron and {Mankovich}, Christopher and {Montgomery}, M.~H. and {Stello}, Dennis and {Timmes}, F.~X. and {Townsend}, Richard},
        title = "{Modules for Experiments in Stellar Astrophysics (MESA): Planets, Oscillations, Rotation, and Massive Stars}",
      journal = {\apjs},
         year = 2013,
        month = sep,
       volume = {208},
       number = {1},
          eid = {4},
        pages = {4},
          doi = {10.1088/0067-0049/208/1/4},
archivePrefix = {arXiv},
       eprint = {1301.0319},
 primaryClass = {astro-ph.SR},
       adsurl = {https://ui.adsabs.harvard.edu/abs/2013ApJS..208....4P}
}

@ARTICLE{2015ApJS..220...15P,
       author = {{Paxton}, Bill and {Marchant}, Pablo and {Schwab}, Josiah and {Bauer}, Evan B. and {Bildsten}, Lars and {Cantiello}, Matteo and {Dessart}, Luc and {Farmer}, R. and {Hu}, H. and {Langer}, N. and {Townsend}, R.~H.~D. and {Townsley}, Dean M. and {Timmes}, F.~X.},
        title = "{Modules for Experiments in Stellar Astrophysics (MESA): Binaries, Pulsations, and Explosions}",
      journal = {\apjs},
         year = 2015,
        month = sep,
       volume = {220},
       number = {1},
          eid = {15},
        pages = {15},
          doi = {10.1088/0067-0049/220/1/15},
archivePrefix = {arXiv},
       eprint = {1506.03146},
 primaryClass = {astro-ph.SR},
       adsurl = {https://ui.adsabs.harvard.edu/abs/2015ApJS..220...15P}
}

@ARTICLE{2018ApJS..234...34P,
       author = {{Paxton}, Bill and {Schwab}, Josiah and {Bauer}, Evan B. and {Bildsten}, Lars and {Blinnikov}, Sergei and {Duffell}, Paul and {Farmer}, R. and {Goldberg}, Jared A. and {Marchant}, Pablo and {Sorokina}, Elena and {Thoul}, Anne and {Townsend}, Richard H.~D. and {Timmes}, F.~X.},
        title = "{Modules for Experiments in Stellar Astrophysics (MESA): Convective Boundaries, Element Diffusion, and Massive Star Explosions}",
      journal = {\apjs},
         year = 2018,
        month = feb,
       volume = {234},
       number = {2},
          eid = {34},
        pages = {34},
          doi = {10.3847/1538-4365/aaa5a8},
archivePrefix = {arXiv},
       eprint = {1710.08424},
 primaryClass = {astro-ph.SR},
       adsurl = {https://ui.adsabs.harvard.edu/abs/2018ApJS..234...34P}
}

@ARTICLE{2019ApJS..243...10P,
       author = {{Paxton}, Bill and {Smolec}, R. and {Schwab}, Josiah and {Gautschy}, A. and {Bildsten}, Lars and {Cantiello}, Matteo and {Dotter}, Aaron and {Farmer}, R. and {Goldberg}, Jared A. and {Jermyn}, Adam S. and {Kanbur}, S.~M. and {Marchant}, Pablo and {Thoul}, Anne and {Townsend}, Richard H.~D. and {Wolf}, William M. and {Zhang}, Michael and {Timmes}, F.~X.},
        title = "{Modules for Experiments in Stellar Astrophysics (MESA): Pulsating Variable Stars, Rotation, Convective Boundaries, and Energy Conservation}",
      journal = {\apjs},
         year = 2019,
        month = jul,
       volume = {243},
       number = {1},
          eid = {10},
        pages = {10},
          doi = {10.3847/1538-4365/ab2241},
archivePrefix = {arXiv},
       eprint = {1903.01426},
 primaryClass = {astro-ph.SR},
       adsurl = {https://ui.adsabs.harvard.edu/abs/2019ApJS..243...10P}
}

@ARTICLE{2023ApJS..265...15J,
       author = {{Jermyn}, Adam S. and {Bauer}, Evan B. and {Schwab}, Josiah and {Farmer}, R. and {Ball}, Warrick H. and {Bellinger}, Earl P. and {Dotter}, Aaron and {Joyce}, Meridith and {Marchant}, Pablo and {Mombarg}, Joey S.~G. and {Wolf}, William M. and {Sunny Wong}, Tin Long and {Cinquegrana}, Giulia C. and {Farrell}, Eoin and {Smolec}, R. and {Thoul}, Anne and {Cantiello}, Matteo and {Herwig}, Falk and {Toloza}, Odette and {Bildsten}, Lars and {Townsend}, Richard H.~D. and {Timmes}, F.~X.},
        title = "{Modules for Experiments in Stellar Astrophysics (MESA): Time-dependent Convection, Energy Conservation, Automatic Differentiation, and Infrastructure}",
      journal = {\apjs},
         year = 2023,
        month = mar,
       volume = {265},
       number = {1},
          eid = {15},
        pages = {15},
          doi = {10.3847/1538-4365/acae8d},
archivePrefix = {arXiv},
       eprint = {2208.03651},
 primaryClass = {astro-ph.SR},
       adsurl = {https://ui.adsabs.harvard.edu/abs/2023ApJS..265...15J}
}

@ARTICLE{1996ApJ...464..943I,
       author = {{Iglesias}, Carlos A. and {Rogers}, Forrest J.},
        title = "{Updated Opal Opacities}",
      journal = {\apj},
         year = 1996,
        month = jun,
       volume = {464},
        pages = {943},
          doi = {10.1086/177381},
       adsurl = {https://ui.adsabs.harvard.edu/abs/1996ApJ...464..943I}
}

@ARTICLE{2005ApJ...623..585F,
       author = {{Ferguson}, Jason W. and {Alexander}, David R. and {Allard}, France and {Barman}, Travis and {Bodnarik}, Julia G. and {Hauschildt}, Peter H. and {Heffner-Wong}, Amanda and {Tamanai}, Akemi},
        title = "{Low-Temperature Opacities}",
      journal = {\apj},
         year = 2005,
        month = apr,
       volume = {623},
       number = {1},
        pages = {585-596},
          doi = {10.1086/428642},
archivePrefix = {arXiv},
       eprint = {astro-ph/0502045},
 primaryClass = {astro-ph},
       adsurl = {https://ui.adsabs.harvard.edu/abs/2005ApJ...623..585F}
}

@ARTICLE{1998SSRv...85..161G,
       author = {{Grevesse}, N. and {Sauval}, A.~J.},
        title = "{Standard Solar Composition}",
      journal = {\ssr},
         year = 1998,
        month = may,
       volume = {85},
        pages = {161-174},
          doi = {10.1023/A:1005161325181},
       adsurl = {https://ui.adsabs.harvard.edu/abs/1998SSRv...85..161G}
}

@ARTICLE{2010ApJS..189..240C,
       author = {{Cyburt}, Richard H. and {Amthor}, A. Matthew and {Ferguson}, Ryan and {Meisel}, Zach and {Smith}, Karl and {Warren}, Scott and {Heger}, Alexander and {Hoffman}, R.~D. and {Rauscher}, Thomas and {Sakharuk}, Alexander and {Schatz}, Hendrik and {Thielemann}, F.~K. and {Wiescher}, Michael},
        title = "{The JINA REACLIB Database: Its Recent Updates and Impact on Type-I X-ray Bursts}",
      journal = {\apjs},
         year = 2010,
        month = jul,
       volume = {189},
       number = {1},
        pages = {240-252},
          doi = {10.1088/0067-0049/189/1/240},
       adsurl = {https://ui.adsabs.harvard.edu/abs/2010ApJS..189..240C}
}

@ARTICLE{2016ApJ...823..102C,
       author = {{Choi}, Jieun and {Dotter}, Aaron and {Conroy}, Charlie and {Cantiello}, Matteo and {Paxton}, Bill and {Johnson}, Benjamin D.},
        title = "{Mesa Isochrones and Stellar Tracks (MIST). I. Solar-scaled Models}",
      journal = {\apj},
         year = 2016,
        month = jun,
       volume = {823},
       number = {2},
          eid = {102},
        pages = {102},
          doi = {10.3847/0004-637X/823/2/102},
archivePrefix = {arXiv},
       eprint = {1604.08592},
 primaryClass = {astro-ph.SR},
       adsurl = {https://ui.adsabs.harvard.edu/abs/2016ApJ...823..102C}
}

@INPROCEEDINGS{2008IAUS..252..205H,
       author = {{Herwig}, Falk},
        title = "{AGB star models. Results from 1D stellar evolution and multi-dimensional hydrodynamics simulations}",
    booktitle = {The Art of Modeling Stars in the 21st Century},
         year = 2008,
       editor = {{Deng}, L. and {Chan}, K.~L.},
       series = {IAU Symposium},
       volume = {252},
        month = oct,
        pages = {205-213},
          doi = {10.1017/S1743921308022795},
       adsurl = {https://ui.adsabs.harvard.edu/abs/2008IAUS..252..205H}
}

@ARTICLE{1975MSRSL...8..369R,
       author = {{Reimers}, D.},
        title = "{Circumstellar absorption lines and mass loss from red giants.}",
      journal = {Memoires of the Societe Royale des Sciences de Liege},
         year = 1975,
        month = jan,
       volume = {8},
        pages = {369-382},
       adsurl = {https://ui.adsabs.harvard.edu/abs/1975MSRSL...8..369R}
}

@ARTICLE{1977A&A....61..217R,
       author = {{Reimers}, D.},
        title = "{On the absolute scale of mass-loss in red giants. I. Circumstellar absorption lines in the spectrum of the visual companion of alpha $^{1}$Her.}",
      journal = {\aap},
         year = 1977,
        month = oct,
       volume = {61},
        pages = {217-224},
       adsurl = {https://ui.adsabs.harvard.edu/abs/1977A&A....61..217R}
}

@ARTICLE{1995A&A...297..727B,
       author = {{Bloecker}, T.},
        title = "{Stellar evolution of low and intermediate-mass stars. I. Mass loss on the AGB and its consequences for stellar evolution.}",
      journal = {\aap},
         year = 1995,
        month = may,
       volume = {297},
        pages = {727},
       adsurl = {https://ui.adsabs.harvard.edu/abs/1995A&A...297..727B}
}

@ARTICLE{2014ApJ...782...17K,
       author = {{Kalirai}, Jason S. and {Marigo}, Paola and {Tremblay}, Pier-Emmanuel},
        title = "{The Core Mass Growth and Stellar Lifetime of Thermally Pulsing Asymptotic Giant Branch Stars}",
      journal = {\apj},
         year = 2014,
        month = feb,
       volume = {782},
       number = {1},
          eid = {17},
        pages = {17},
          doi = {10.1088/0004-637X/782/1/17},
archivePrefix = {arXiv},
       eprint = {1312.4544},
 primaryClass = {astro-ph.SR},
       adsurl = {https://ui.adsabs.harvard.edu/abs/2014ApJ...782...17K}
}

@ARTICLE{2012A&A...542A...1L,
       author = {{Lau}, H.~H.~B. and {Gil-Pons}, P. and {Doherty}, C. and {Lattanzio}, J.},
        title = "{The end of super AGB and massive AGB stars. I. The instabilities that determine the final mass of AGB stars}",
      journal = {\aap},
         year = 2012,
        month = jun,
       volume = {542},
          eid = {A1},
        pages = {A1},
          doi = {10.1051/0004-6361/201218826},
archivePrefix = {arXiv},
       eprint = {1204.3931},
 primaryClass = {astro-ph.SR},
       adsurl = {https://ui.adsabs.harvard.edu/abs/2012A&A...542A...1L}
}

@ARTICLE{2024MNRAS.531.4033R,
       author = {{Rees}, Natalie R. and {Izzard}, Robert G.},
        title = "{Evolving past instabilities on the thermally pulsing-(super)asymptotic giant branch}",
      journal = {\mnras},
         year = 2024,
        month = jul,
       volume = {531},
       number = {4},
        pages = {4033-4044},
          doi = {10.1093/mnras/stae1387},
       adsurl = {https://ui.adsabs.harvard.edu/abs/2024MNRAS.531.4033R}
}

@ARTICLE{1979ApJS...41..513M,
       author = {{Miller}, G.~E. and {Scalo}, J.~M.},
        title = "{The Initial Mass Function and Stellar Birthrate in the Solar Neighborhood}",
      journal = {\apjs},
         year = 1979,
        month = nov,
       volume = {41},
        pages = {513},
          doi = {10.1086/190629},
       adsurl = {https://ui.adsabs.harvard.edu/abs/1979ApJS...41..513M}
}

@ARTICLE{2020RAA....20..161H,
       author = {{Han}, Zhan-Wen and {Ge}, Hong-Wei and {Chen}, Xue-Fei and {Chen}, Hai-Liang},
        title = "{Binary Population Synthesis}",
      journal = {Research in Astronomy and Astrophysics},
         year = 2020,
        month = oct,
       volume = {20},
       number = {10},
          eid = {161},
        pages = {161},
          doi = {10.1088/1674-4527/20/10/161},
archivePrefix = {arXiv},
       eprint = {2009.08611},
 primaryClass = {astro-ph.SR},
       adsurl = {https://ui.adsabs.harvard.edu/abs/2020RAA....20..161H}
}

@ARTICLE{1944MNRAS.104..273B,
       author = {{Bondi}, H. and {Hoyle}, F.},
        title = "{On the mechanism of accretion by stars}",
      journal = {\mnras},
         year = 1944,
        month = jan,
       volume = {104},
        pages = {273},
          doi = {10.1093/mnras/104.5.273},
       adsurl = {https://ui.adsabs.harvard.edu/abs/1944MNRAS.104..273B}
}

@ARTICLE{1988A&A...205..155B,
       author = {{Boffin}, H.~M.~J. and {Jorissen}, A.},
        title = "{Can a barium star be produced by wind accretion in a detached binary ?}",
      journal = {\aap},
         year = 1988,
        month = oct,
       volume = {205},
        pages = {155-163},
       adsurl = {https://ui.adsabs.harvard.edu/abs/1988A&A...205..155B}
}

@ARTICLE{1997MNRAS.291..732T,
       author = {{Tout}, Christopher A. and {Aarseth}, Sverre J. and {Pols}, Onno R. and {Eggleton}, Peter P.},
        title = "{Rapid binary star evolution for N-body simulations and population synthesis}",
      journal = {\mnras},
         year = 1997,
        month = nov,
       volume = {291},
       number = {4},
        pages = {732-748},
          doi = {10.1093/mnras/291.4.732},
       adsurl = {https://ui.adsabs.harvard.edu/abs/1997MNRAS.291..732T}
}

@ARTICLE{2007MNRAS.376...61D,
       author = {{Dray}, Lynnette M. and {Tout}, Christopher A.},
        title = "{On rejuvenation in massive binary systems}",
      journal = {\mnras},
         year = 2007,
        month = mar,
       volume = {376},
       number = {1},
        pages = {61-70},
          doi = {10.1111/j.1365-2966.2007.11431.x},
archivePrefix = {arXiv},
       eprint = {astro-ph/0612539},
 primaryClass = {astro-ph},
       adsurl = {https://ui.adsabs.harvard.edu/abs/2007MNRAS.376...61D}
}

@BOOK{1992adps.book.....L,
       author = {{Lang}, Kenneth R.},
        title = "{Astrophysical Data I. Planets and Stars.}",
         year = 1992,
    publisher = {Springer-Verlag},
       adsurl = {https://ui.adsabs.harvard.edu/abs/1992adps.book.....L}
}

@ARTICLE{1981A&A....99..126H,
       author = {{Hut}, P.},
        title = "{Tidal evolution in close binary systems.}",
      journal = {\aap},
         year = 1981,
        month = jun,
       volume = {99},
        pages = {126-140},
       adsurl = {https://ui.adsabs.harvard.edu/abs/1981A&A....99..126H}
}

@ARTICLE{1975A&A....41..329Z,
       author = {{Zahn}, J.-P.},
        title = "{The dynamical tide in close binaries.}",
      journal = {\aap},
         year = 1975,
        month = jul,
       volume = {41},
        pages = {329-344},
       adsurl = {https://ui.adsabs.harvard.edu/abs/1975A&A....41..329Z}
}

@ARTICLE{1977A&A....57..383Z,
       author = {{Zahn}, J.-P.},
        title = "{Tidal friction in close binary systems.}",
      journal = {\aap},
         year = 1977,
        month = may,
       volume = {57},
        pages = {383-394},
       adsurl = {https://ui.adsabs.harvard.edu/abs/1977A&A....57..383Z}
}

@ARTICLE{1988ApJ...329..764L,
       author = {{Livio}, Mario and {Soker}, Noam},
        title = "{The Common Envelope Phase in the Evolution of Binary Stars}",
      journal = {\apj},
         year = 1988,
        month = jun,
       volume = {329},
        pages = {764},
          doi = {10.1086/166419},
       adsurl = {https://ui.adsabs.harvard.edu/abs/1988ApJ...329..764L}
}

@ARTICLE{2010ApJ...716..114X,
       author = {{Xu}, Xiao-Jie and {Li}, Xiang-Dong},
        title = "{On the Binding Energy Parameter {\ensuremath{\lambda}} of Common Envelope Evolution}",
      journal = {\apj},
         year = 2010,
        month = jun,
       volume = {716},
       number = {1},
        pages = {114-121},
          doi = {10.1088/0004-637X/716/1/114},
archivePrefix = {arXiv},
       eprint = {1004.4957},
 primaryClass = {astro-ph.SR},
       adsurl = {https://ui.adsabs.harvard.edu/abs/2010ApJ...716..114X}
}

@ARTICLE{2010ApJ...722.1985X,
       author = {{Xu}, Xiao-Jie and {Li}, Xiang-Dong},
        title = "{Erratum: ``On the Binding Energy Parameter {\ensuremath{\lambda}} of Common Envelope Evolution'' (2010, ApJ, 716, 114)}",
      journal = {\apj},
         year = 2010,
        month = oct,
       volume = {722},
       number = {2},
        pages = {1985-1988},
          doi = {10.1088/0004-637X/722/2/1985},
       adsurl = {https://ui.adsabs.harvard.edu/abs/2010ApJ...722.1985X}
}

@ARTICLE{2011ApJ...743...49L,
       author = {{Loveridge}, A.~J. and {van der Sluys}, M.~V. and {Kalogera}, V.},
        title = "{Analytical Expressions for the Envelope Binding Energy of Giants as a Function of Basic Stellar Parameters}",
      journal = {\apj},
         year = 2011,
        month = dec,
       volume = {743},
       number = {1},
          eid = {49},
        pages = {49},
          doi = {10.1088/0004-637X/743/1/49},
archivePrefix = {arXiv},
       eprint = {1009.5400},
 primaryClass = {astro-ph.SR},
       adsurl = {https://ui.adsabs.harvard.edu/abs/2011ApJ...743...49L}
}

@ARTICLE{2016RAA....16..126W,
       author = {{Wang}, Chen and {Jia}, Kun and {Li}, Xiang-Dong},
        title = "{The binding energy parameter for common envelope evolution}",
      journal = {Research in Astronomy and Astrophysics},
         year = 2016,
        month = aug,
       volume = {16},
       number = {8},
          eid = {126},
        pages = {126},
          doi = {10.1088/1674-4527/16/8/126},
archivePrefix = {arXiv},
       eprint = {1605.03668},
 primaryClass = {astro-ph.SR},
       adsurl = {https://ui.adsabs.harvard.edu/abs/2016RAA....16..126W}
}

@ARTICLE{2016MNRAS.463.1040M,
       author = {{Madappatt}, Niyas and {De Marco}, Orsola and {Villaver}, Eva},
        title = "{The effect of tides on the population of PN from interacting binaries}",
      journal = {\mnras},
         year = 2016,
        month = nov,
       volume = {463},
       number = {1},
        pages = {1040-1056},
          doi = {10.1093/mnras/stw2025},
archivePrefix = {arXiv},
       eprint = {1608.03041},
 primaryClass = {astro-ph.SR},
       adsurl = {https://ui.adsabs.harvard.edu/abs/2016MNRAS.463.1040M}
}

@ARTICLE{2014A&A...568A..68Z,
       author = {{Zorotovic}, M. and {Schreiber}, M.~R. and {Garc{\'\i}a-Berro}, E. and {Camacho}, J. and {Torres}, S. and {Rebassa-Mansergas}, A. and {G{\"a}nsicke}, B.~T.},
        title = "{Monte Carlo simulations of post-common-envelope white dwarf + main sequence binaries: The effects of including recombination energy}",
      journal = {\aap},
         year = 2014,
        month = aug,
       volume = {568},
          eid = {A68},
        pages = {A68},
          doi = {10.1051/0004-6361/201323039},
archivePrefix = {arXiv},
       eprint = {1407.3301},
 primaryClass = {astro-ph.SR},
       adsurl = {https://ui.adsabs.harvard.edu/abs/2014A&A...568A..68Z}
}

@ARTICLE{2024A&A...682A..33B,
       author = {{Belloni}, Diogo and {Schreiber}, Matthias R. and {Moe}, Maxwell and {El-Badry}, Kareem and {Shen}, Ken J.},
        title = "{Evidence for saturated and disrupted magnetic braking from samples of detached close binaries with M and K dwarfs}",
      journal = {\aap},
         year = 2024,
        month = feb,
       volume = {682},
          eid = {A33},
        pages = {A33},
          doi = {10.1051/0004-6361/202347931},
archivePrefix = {arXiv},
       eprint = {2311.04309},
 primaryClass = {astro-ph.SR},
       adsurl = {https://ui.adsabs.harvard.edu/abs/2024A&A...682A..33B}
}

@ARTICLE{2024PASP..136l4201B,
       author = {{Blomberg}, Lisa and {El-Badry}, Kareem and {Breivik}, Katelyn and {Caiazzo}, Ilaria and {Nagarajan}, Pranav and {Rodriguez}, Antonio and {van Roestel}, Jan and {Vanderbosch}, Zachary P. and {Yamaguchi}, Natsuko},
        title = "{The Companion Mass Distribution of Post Common Envelope Hot Subdwarf Binaries: Evidence for Boosted and Disrupted Magnetic Braking?}",
      journal = {\pasp},
         year = 2024,
        month = dec,
       volume = {136},
       number = {12},
          eid = {124201},
        pages = {124201},
          doi = {10.1088/1538-3873/ad94a2},
archivePrefix = {arXiv},
       eprint = {2408.15334},
 primaryClass = {astro-ph.SR},
       adsurl = {https://ui.adsabs.harvard.edu/abs/2024PASP..136l4201B}
}

@ARTICLE{2024ApJ...970L..11H,
       author = {{Hallakoun}, Na'ama and {Shahaf}, Sahar and {Mazeh}, Tsevi and {Toonen}, Silvia and {Ben-Ami}, Sagi},
        title = "{A Deficit of Massive White Dwarfs in Gaia Astrometric Binaries}",
      journal = {\apjl},
         year = 2024,
        month = jul,
       volume = {970},
       number = {1},
          eid = {L11},
        pages = {L11},
          doi = {10.3847/2041-8213/ad5e63},
archivePrefix = {arXiv},
       eprint = {2311.17145},
 primaryClass = {astro-ph.SR},
       adsurl = {https://ui.adsabs.harvard.edu/abs/2024ApJ...970L..11H}
}

@ARTICLE{2026ApJ..1010...73T,
       author = {{Thai}, Riley and {Karakas}, Amanda I. and {Osborn}, Zara and {Izzard}, Robert G. and {Hirai}, Ryosuke and {Kemp}, Alex J. and {Campbell}, Simon W.},
        title = "{The Best Guess: Testing New and Old Formalisms for the Common Envelope against Observations}",
      journal = {\apj},
         year = 2026,
        month = oct,
       volume = {1010},
       number = {1},
          eid = {73},
        pages = {73},
          doi = {10.3847/1538-4357/ae9c3c},
       adsurl = {https://ui.adsabs.harvard.edu/abs/2026ApJ..1010...73T}
}

@ARTICLE{2020ApJS..249....9G,
       author = {{Ge}, Hongwei and {Webbink}, Ronald F. and {Han}, Zhanwen},
        title = "{The Thermal Equilibrium Mass-loss Model and Its Applications in Binary Evolution}",
      journal = {\apjs},
         year = 2020,
        month = jul,
       volume = {249},
       number = {1},
          eid = {9},
        pages = {9},
          doi = {10.3847/1538-4365/ab98f6},
archivePrefix = {arXiv},
       eprint = {2006.00774},
 primaryClass = {astro-ph.SR},
       adsurl = {https://ui.adsabs.harvard.edu/abs/2020ApJS..249....9G}
}

@ARTICLE{2026A&A...707A.342C,
       author = {{Cui}, Yingzhen and {Wang}, Song and {Meng}, Xiangcun and {Liu}, Jifeng and {Ma}, Shuguo and {Zhao}, Weitao},
        title = "{Dynamical mass loss at the end of thermally pulsating asymptotic giant branch stars}",
      journal = {\aap},
         year = 2026,
        month = mar,
       volume = {707},
          eid = {A342},
        pages = {A342},
          doi = {10.1051/0004-6361/202556962},
archivePrefix = {arXiv},
       eprint = {2601.18194},
 primaryClass = {astro-ph.SR},
       adsurl = {https://ui.adsabs.harvard.edu/abs/2026A&A...707A.342C}
}

@ARTICLE{2010ApJ...717L..62Y,
       author = {{Yoon}, Sung-Chul and {Cantiello}, Matteo},
        title = "{Evolution of Massive Stars with Pulsation-driven Superwinds During the Red Supergiant Phase}",
      journal = {\apjl},
         year = 2010,
        month = jul,
       volume = {717},
       number = {1},
        pages = {L62-L65},
          doi = {10.1088/2041-8205/717/1/L62},
archivePrefix = {arXiv},
       eprint = {1005.4925},
 primaryClass = {astro-ph.SR},
       adsurl = {https://ui.adsabs.harvard.edu/abs/2010ApJ...717L..62Y}
}

@ARTICLE{2010A&A...520A..86Z,
       author = {{Zorotovic}, M. and {Schreiber}, M.~R. and {G{\"a}nsicke}, B.~T. and {Nebot G{\'o}mez-Mor{\'a}n}, A.},
        title = "{Post-common-envelope binaries from SDSS. IX: Constraining the common-envelope efficiency}",
      journal = {\aap},
         year = 2010,
        month = sep,
       volume = {520},
          eid = {A86},
        pages = {A86},
          doi = {10.1051/0004-6361/200913658},
archivePrefix = {arXiv},
       eprint = {1006.1621},
 primaryClass = {astro-ph.SR},
       adsurl = {https://ui.adsabs.harvard.edu/abs/2010A&A...520A..86Z}
}

@ARTICLE{2013A&A...557A..87T,
       author = {{Toonen}, S. and {Nelemans}, G.},
        title = "{The effect of common-envelope evolution on the visible population of post-common-envelope binaries}",
      journal = {\aap},
         year = 2013,
        month = sep,
       volume = {557},
          eid = {A87},
        pages = {A87},
          doi = {10.1051/0004-6361/201321753},
archivePrefix = {arXiv},
       eprint = {1309.0327},
 primaryClass = {astro-ph.SR},
       adsurl = {https://ui.adsabs.harvard.edu/abs/2013A&A...557A..87T}
}

@ARTICLE{2017MNRAS.470.1442C,
       author = {{Cojocaru}, R. and {Rebassa-Mansergas}, A. and {Torres}, S. and {Garc{\'\i}a-Berro}, E.},
        title = "{The population of white dwarf-main sequence binaries in the SDSS DR 12}",
      journal = {\mnras},
         year = 2017,
        month = sep,
       volume = {470},
       number = {2},
        pages = {1442-1452},
          doi = {10.1093/mnras/stx1326},
archivePrefix = {arXiv},
       eprint = {1705.05888},
 primaryClass = {astro-ph.SR},
       adsurl = {https://ui.adsabs.harvard.edu/abs/2017MNRAS.470.1442C}
}

@ARTICLE{2022ApJ...933..137G,
       author = {{Ge}, Hongwei and {Tout}, Christopher A. and {Chen}, Xuefei and {Kruckow}, Matthias U. and {Chen}, Hailiang and {Jiang}, Dengkai and {Li}, Zhenwei and {Liu}, Zhengwei and {Han}, Zhanwen},
        title = "{The Common Envelope Evolution Outcome-A Case Study on Hot Subdwarf B Stars}",
      journal = {\apj},
         year = 2022,
        month = jul,
       volume = {933},
       number = {2},
          eid = {137},
        pages = {137},
          doi = {10.3847/1538-4357/ac75d3},
archivePrefix = {arXiv},
       eprint = {2205.14256},
 primaryClass = {astro-ph.SR},
       adsurl = {https://ui.adsabs.harvard.edu/abs/2022ApJ...933..137G}
}

@ARTICLE{2024ApJ...961..202G,
       author = {{Ge}, Hongwei and {Tout}, Christopher A. and {Webbink}, Ronald F. and {Chen}, Xuefei and {Sarkar}, Arnab and {Li}, Jiao and {Li}, Zhenwei and {Zhang}, Lifu and {Han}, Zhanwen},
        title = "{The Common Envelope Evolution Outcome. II. Short-orbital-period Hot Subdwarf B Binaries Reveal a Clear Picture}",
      journal = {\apj},
         year = 2024,
        month = feb,
       volume = {961},
       number = {2},
          eid = {202},
        pages = {202},
          doi = {10.3847/1538-4357/ad158e},
archivePrefix = {arXiv},
       eprint = {2311.17304},
 primaryClass = {astro-ph.SR},
       adsurl = {https://ui.adsabs.harvard.edu/abs/2024ApJ...961..202G}
}

@ARTICLE{2025A&A...695A.161S,
       author = {{Santos-Garc{\'\i}a}, A. and {Torres}, S. and {Rebassa-Mansergas}, A. and {Brown}, A.~J.},
        title = "{A population synthesis study of the Gaia 100 pc unresolved white dwarf-main-sequence binary population}",
      journal = {\aap},
         year = 2025,
        month = mar,
       volume = {695},
          eid = {A161},
        pages = {A161},
          doi = {10.1051/0004-6361/202452989},
archivePrefix = {arXiv},
       eprint = {2502.11593},
 primaryClass = {astro-ph.SR},
       adsurl = {https://ui.adsabs.harvard.edu/abs/2025A&A...695A.161S}
}

@ARTICLE{2018MNRAS.478.1818G,
       author = {{Grichener}, Aldana and {Sabach}, Efrat and {Soker}, Noam},
        title = "{The limited role of recombination energy in common envelope removal}",
      journal = {\mnras},
         year = 2018,
        month = aug,
       volume = {478},
       number = {2},
        pages = {1818-1824},
          doi = {10.1093/mnras/sty1178},
archivePrefix = {arXiv},
       eprint = {1803.05864},
 primaryClass = {astro-ph.SR},
       adsurl = {https://ui.adsabs.harvard.edu/abs/2018MNRAS.478.1818G}
}

@ARTICLE{2022MNRAS.516.4669L,
       author = {{Lau}, Mike Y.~M. and {Hirai}, Ryosuke and {Price}, Daniel J. and {Mandel}, Ilya},
        title = "{Common envelopes in massive stars II: The distinct roles of hydrogen and helium recombination}",
      journal = {\mnras},
         year = 2022,
        month = nov,
       volume = {516},
       number = {4},
        pages = {4669-4678},
          doi = {10.1093/mnras/stac2490},
archivePrefix = {arXiv},
       eprint = {2206.06411},
 primaryClass = {astro-ph.SR},
       adsurl = {https://ui.adsabs.harvard.edu/abs/2022MNRAS.516.4669L}
}

@ARTICLE{2024MNRAS.528..234C,
       author = {{Chamandy}, Luke and {Carroll-Nellenback}, Jonathan and {Blackman}, Eric G. and {Frank}, Adam and {Tu}, Yisheng and {Liu}, Baowei and {Zou}, Yangyuxin and {Nordhaus}, Jason},
        title = "{How negative feedback and the ambient environment limit the influence of recombination in common envelope evolution}",
      journal = {\mnras},
         year = 2024,
        month = feb,
       volume = {528},
       number = {1},
        pages = {234-254},
          doi = {10.1093/mnras/stae036},
archivePrefix = {arXiv},
       eprint = {2304.14840},
 primaryClass = {astro-ph.SR},
       adsurl = {https://ui.adsabs.harvard.edu/abs/2024MNRAS.528..234C}
}

@ARTICLE{2026arXiv260720611G,
       author = {{Grondin}, Steffani M. and {Drout}, Maria R. and {Nordhaus}, Jason and {Muirhead}, Philip S. and {Filer}, Bailey and {Laroche}, Alexander and {Webb}, Jeremy J. and {Broekgaarden}, Floor S. and {Chornock}, Ryan and {Kremer}, Kyle and {LeBaron}, Natalie and {Margutti}, Raffaella and {Noughani}, Nikki and {Sears}, Huei and {Tremblay}, Pier-Emmanuel},
        title = "{A Framework for Linking Pre- and Post-Common Envelope Binary Properties with Star Clusters: The First Demonstration with a Massive White Dwarf+M Dwarf Binary in Alessi 12}",
      journal = {arXiv e-prints},
         year = 2026,
        month = jul,
          eid = {arXiv:2607.20611},
        pages = {arXiv:2607.20611},
          doi = {10.48550/arXiv.2607.20611},
archivePrefix = {arXiv},
       eprint = {2607.20611},
 primaryClass = {astro-ph.SR},
       adsurl = {https://ui.adsabs.harvard.edu/abs/2026arXiv260720611G}
}

@ARTICLE{2019ApJ...880...33Y,
       author = {{Yahalomi}, Daniel A. and {Shvartzvald}, Yossi and {Agol}, Eric and {Shporer}, Avi and {Latham}, David W. and {Kruse}, Ethan and {Brewer}, John M. and {Buchhave}, Lars A. and {Fulton}, Benjamin J. and {Howard}, Andrew W. and {Isaacson}, Howard and {Petigura}, Erik A. and {Quinn}, Samuel N.},
        title = "{The Mass of the White Dwarf Companion in the Self-lensing Binary KOI-3278: Einstein versus Newton}",
      journal = {\apj},
         year = 2019,
        month = jul,
       volume = {880},
       number = {1},
          eid = {33},
        pages = {33},
          doi = {10.3847/1538-4357/ab2649},
archivePrefix = {arXiv},
       eprint = {1904.11063},
 primaryClass = {astro-ph.SR}
}

@ARTICLE{2014Sci...344..275K,
       author = {{Kruse}, Ethan and {Agol}, Eric},
        title = "{KOI-3278: A Self-Lensing Binary Star System}",
      journal = {Science},
         year = 2014,
        month = apr,
       volume = {344},
       number = {6181},
        pages = {275-277},
          doi = {10.1126/science.1251999},
       adsurl = {https://ui.adsabs.harvard.edu/abs/2014Sci...344..275K}
}

@ARTICLE{2018AJ....155..144K,
       author = {{Kawahara}, Hajime and {Masuda}, Kento and {MacLeod}, Morgan and {Latham}, David W. and {Bieryla}, Allyson and {Benomar}, Othman},
        title = "{Discovery of Three Self-lensing Binaries from Kepler}",
      journal = {\aj},
         year = 2018,
        month = mar,
       volume = {155},
       number = {3},
          eid = {144},
        pages = {144},
          doi = {10.3847/1538-3881/aaaaaf},
archivePrefix = {arXiv},
       eprint = {1801.07874},
 primaryClass = {astro-ph.SR},
       adsurl = {https://ui.adsabs.harvard.edu/abs/2018AJ....155..144K}
}
\bibliographystyle{aasjournalv7.1}

\end{document}